\documentclass[lettersize,journal]{IEEEtran}
\usepackage{amsmath,amsfonts}
\usepackage{algorithmic}
\usepackage{algorithm}
\usepackage{array}
\usepackage[caption=false,font=normalsize,labelfont=sf,textfont=sf]{subfig}
\usepackage{textcomp}
\usepackage{stfloats}
\usepackage{url}
\usepackage{verbatim}
\usepackage{graphicx}
\usepackage{cite}
\begin{document}

\title{High-Resolution WiFi Imaging with Reconfigurable Intelligent Surfaces}

\author{Ying He, Dongheng Zhang,~\IEEEmembership{Member,~IEEE,} Yan Chen,~\IEEEmembership{Senior Member,~IEEE,}
	\thanks{Ying He is with the School of Information and Communication, University of Electronic Science and Technology of China, Chengdu 611731, China (e-mail: heying@std.uestc.edu.cn).}
	\thanks{Dongheng Zhang and Yan Chen are with School of Cyber Science and Technology, University of Science and Technology of China, Hefei 230026, China (email: dongheng@ustc.edu.cn; eecyan@ustc.edu.cn)}
	\thanks{}}

\markboth{}%
{Shell \MakeLowercase{\textit{et al.}}: A Sample Article Using IEEEtran.cls for IEEE Journals}

\IEEEpubid{}

\maketitle

\begin{abstract}
	WiFi-based imaging enables pervasive sensing in a privacy-preserving and cost-effective way. 
	However, most of existing methods either require specialized hardware modification or suffer from poor imaging performance due to the fundamental limit of off-the-shelf commodity WiFi devices in spatial resolution.
	We observe that the recently developed reconfigurable intelligent surface (RIS) could be a promising solution to overcome these challenges. Thus, in this paper, we propose a RIS-aided WiFi imaging framework to achieve high-resolution imaging with the off-the-shelf WiFi devices. Specifically, we first design a beamforming method to achieve the first-stage imaging by separating the signals from different spatial locations with the aid of the RIS. Then, we propose an optimization-based super-resolution imaging algorithm by leveraging the low rank nature of the reconstructed object. During the optimization, we also explicitly take into account the effect of finite phase quantization in RIS to avoid the resolution degradation due to quantization errors. Simulation results demonstrate that our framework achieves median root mean square error (RMSE) of 0.03 and median structural similarity (SSIM) of 0.52. The visual results show that high-resolution imaging results are achieved with simulation signals at 5 GHz that are matched with commercial WiFi 802.11n/ac protocols. 
\end{abstract}

\begin{IEEEkeywords}
	WiFi Imaging, Reconfigurable Intelligent Surfaces, Beamforming, Low-rank
\end{IEEEkeywords}

\section{Introduction}

\IEEEPARstart{O}{bject} imaging has always been an important topic in academic and industrial communities. With the information conveyed by image, pervasive human and object sensing with rich contexts (not only the location but also silhouette/shape, size, pose, etc.) can be realized, which enables new applications including human-computer interaction, ubiquitous computing and surveillance.
Although visible light is the most common media for imaging, it suffers from severe quality degradation under poor lighting conditions and blockage of obstacles~\cite{8567956,8957313,9198891,9137344,9373293,ahn2019convolutional,8084714,9200682}. 
To resolve this challenge, Radio Frequency (RF) signal, which is independent of light conditions and can traverse obstacles, has been an alternative for imaging\cite{wang2017tagscan,9187251,2020MTrack,7444139,adib2015capturing,9016250,9234049,8909373,9316693,9442375,8338430}. 
Among several types of RF-based imaging methods, WiFi-based ones have attracted significant interests given the ubiquity of WiFi devices. 

Despite of the advances in reusing WiFi devices for wireless sensing, such as tracking and localization\cite{6555311,8688470,8798719,zhang2019calibrating}, activity sensing and recognition\cite{9205901,9076313,7803604,2017TRIEDS}, vital sign monitoring\cite{9414926,8612907}, etc., WiFi imaging still faces significant challenges and remains unsolved.  
Wision \cite{huang2014feasibility} is a pioneering study which evaluates the feasibility of WiFi imaging. By emulating a ($8\times8 $) antenna array based on the idea of Synthetic Aperture Radar (SAR), it can generate a bubble-like 2D heatmap to image a single static object. With a similar idea built upon SAR, Karanam et al. \cite{7944785} profile the shape of the single-type object more accurately by associating the WiFi Received Signal Strength Indicator~(RSSI). The system utilizes unmanned vehicles to physically coordinate and scan an area to form a virtual antenna array with $150\times150$ elements. Holl et al.\cite{holl2017holography} is the first work that utilizes scattering principles and microwave imaging theorems for modeling the imaging problem. However, this technique requires measurements to be captured in a phase-coherent fashion using two antennas, one is fixed as a reference antenna and another is moved on a 2D plane for scanning. These prior works use 2.4GHz/5GHz WiFi devices, which can only extend the effective aperture by SAR to break their limitation of antenna aperture \cite{huang2014feasibility,holl2017holography,vakalis2019imaging,7944785 }. This leads to the fact that imaging quality is vulnerable under the moving trajectory tracking error and makes it impossible to achieve real-time imaging. 

Although aforementioned methods have achieved promising results, they all require the WiFi antennas to move with given trajectory to form virtual antennas, which is time-consuming and not practical in deployment. The core limitation of these methods lies in the fact that the spatial resolution of WiFi signal is limited by the antenna aperture and signal bandwidth. 
Fortunately, reconfigurable intelligent surfaces (RISs) were recently developed to customize the radio environment, which becomes a promising solution to address this problem \cite{9456027,9086766}.
An RIS is generally a man-made surface composed of a large number of passive elements, each of which is able to induce a controllable amplitude and/or phase change to the incident signal independently\cite{wu_intelligent_2021}.  
By tuning the phase shifts of all elements according to the measurement of wireless channels, the signals reflected by an RIS can superimpose constructively or destructively to achieve fine-grained passive beamforming for directional signal enhancement or nulling.
This thus provides a new degree of freedom to break through the fundamental limitation of WiFi-based imaging performance and paves the way to realize a smart and programmable wireless environment\cite{wu_towards_2020}. 

Compared with existing technologies, RIS possesses various advantages from three aspects.
First, compared with active RF devices, RIS with only passive reflecting elements can potentially yield superior performance with the increasing number of elements while implemented with orders-of-magnitude lower hardware/energy cost \cite{8319526}.
Second, RIS can be intergrated with cellular and WiFi systems while does not require any modification for existing devices. 
Finally, RIS operates in full-duplex (FD) mode without any antenna noise amplification as well as self-interference, which thus offers competitive advantages over traditional active relays, e.g., half-duplex (HD) relay which suffers from low spectral efficiency as well as FD relay which needs sophisticated techniques for self-interference cancellation \cite{wu_intelligent_2021}.
Benefiting from the aforementioned advantages of RIS, RIS-aided wireless communication has attracted a lot of interest in the recent decade and shown its superiority for applications including physical layer security\cite{9264659,8972400,8847342}, wireless power transfer \cite{9133435}, interference suppression\cite{9090356,9279253} and coverage extension \cite{wu_towards_2020}.

In this paper, we present a high-resolution WiFi imaging system with the assistance of RIS. To the best of our knowledge, this paper offers the first contribution aimed at realizing  RIS-aided WiFi imaging. We consider a scenario where two commodity WiFi devices act as the transmitter and the receiver, while an RIS is utilized to assist the WiFi receiver for imaging the target in the area of interest. To achieve this, we have noted that the performance of conventional WiFi imaging is limited by the antenna aperture, which however could be dramatically increased with RIS. In traditional WiFi-based sensing frameworks [cite related reference], the signal intensity from a specific direction and distance is derived by adjusting the phase shift on different antennas and subcarriers. Inspired by these works, we adjust the reflected signal phase on different RIS elements to beamform the echoes to specific position. 
With the existence of echoes in a position, the echoes reflected by RIS would superpose coherently, which leads to a large amplitude in the received signal. Different from leveraging tens of active antennas to generate sharp directional beams, an RIS-aided imaging system allows the APs to be equipped with substantially less antennas while exploiting the large aperture of RIS to create fine-grained reflect beams. In this way, the hardware cost and energy consumption of the system can be significantly reduced. 

\begin{figure*}[t]
	\centering
	\includegraphics[scale = 0.5]{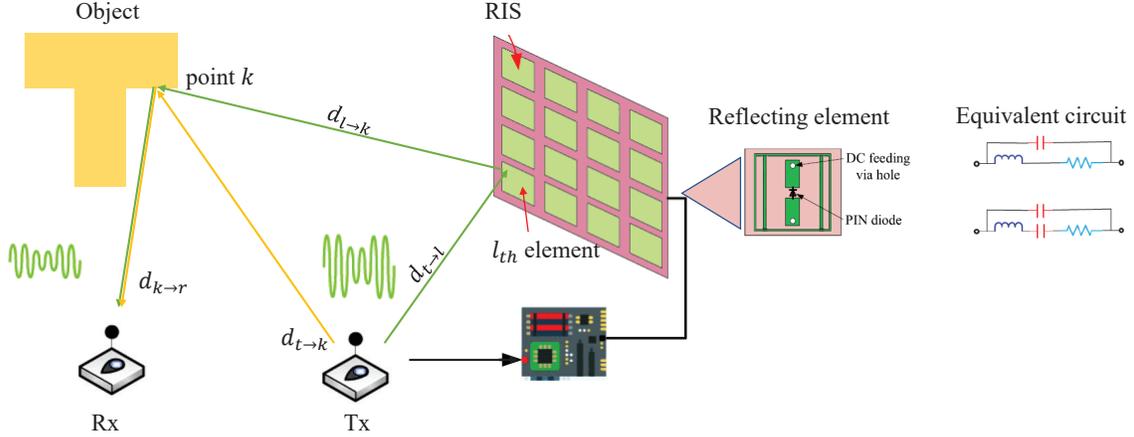}
	\caption{An illustration of the proposed RIS-aided WiFi imaging system. The system contains two single-antenna APs as the transmitter and the receiver, and a 2-D RIS to provide assistance for the imaging. }
	\label{system}
\end{figure*}

However, the resolution of WiFi imaging with only RIS-assisted beamforming is still limited due to two facts. Firstly, the spatial resolution of RF-based image is not comparable with camera-based photos even with a large antenna array due to the relatively large wavelangth of RF signals. Secondly, the WiFi imaging obtained by beamforming is sensitive to phase-shift error, which could not be avoided in existing RISs since they could only achieve finite number of phase shift values. Specifically, the phase shift control in existing system is mainly achieved using positive-intrinsic-negative (PIN) diodes. In Fig.~\ref{system}, we show an example of a design of the reflecting element and its equivalent circuit based on the PIN diode. By applying different biasing voltages to the PIN, the element can achieve a binary phase shifting\cite{wu_towards_2020}. Thus, more phase-shift levels would require more PIN diodes and result in a higher cost. As such, for practical RISs with a large number of elements, it is more cost-effective to implement only discrete phase shifts with a small number of control bits for each element, e.g., 1-bit for two-level (0 or $\pi$) phase shifts\cite{8683145}.

To address these issues, we have noted that the objects to be imaged generally have low rank nature, which has not been exploited as a prior knowledge. Based on this observation, we propose an optimization-based method to enhance the resolution of the imaging results. Specifically, we first design an end-to-end mapping operator between the beamforming output and the object's reflection coefficient (albedo), which is utilized to reformulate the imaging process more precisely. Then, we formulate an optimization problem, of which the constraint is the rank of imaging results. With such a design, the resolution of the imaging results could be enhanced remarkably using the low rank prior. We validate the performance of the proposed optimization-based algorithm with two single-antenna commodity WiFi devices and an RIS of $16\times16$ elements. The simulation results of different objects demonstrate that our algorithm achieves accurate imaging results, visually close to the ground truth, as shown in Fig.~3, with a median root mean square error (RMSE) of 0.03 and a median structural similarity (SSIM) of 0.52. 

The rest of this paper is organized as follows. We describe the system model in Section II. Section III introduces the beamforming-based reconstruction algorithm. The proposed optimization-based algorithm is introduced in Section IV. The simulation results are illustrated and discussed in Section V. Finally the conclusions are drawn in Section VI. 

\section{System model}
The system model of the proposed RIS-assisted WiFi imaging system is shown in Fig.~\ref{system}, which is composed of two single-antenna WiFi Access Points (APs), an RIS with $M\times N$ reflecting elements arranged in a uniform rectangular array (URA), and an object to be imaged. 
Two APs are placed in the horizontal plane along the $y$-axis direction which act as the transmitter and the receiver, respectively. 
The RIS is placed in the vertical plane, where each RIS is connected to a smart controller for adjusting the phase shifts of reflecting elements by adding biasing voltage on PIN diodes. 

In the RIS-aided imaging model, the propagation paths can be divided into two categories according to whether the signal is reflected by the RIS. 
One is the direct path denoted by the orange arrow in Fig.~\ref{system}, of which the signals are transmitted to the object and then reflected back to the receiver directly.
The other one is the RIS path which is denoted by the green arrow in Fig.~\ref{system}, which is the superposition of three components: 
\begin{itemize}
\item $d_{t \rightarrow l}$ : the path from the transmitter to the RIS. 
\item $d_{l \rightarrow k}$ : the reflection from the RIS to the object. 
\item $d_{k \rightarrow r}$ : the path from the object to the receiver. 
\end{itemize}
The length of the direct path passed by point $k$ is denoted as $d_{de}(k) = d_{t k}+d_{r k}$, and the length of the RIS path is denoted as $d_{ris,l}(k) = d_{t l}+d_{l k}+d_{r k}$, where $d_{t k}, d_{r k}, d_{t l}$, and $d_{l k} $ are the distance from the Tx to the object point $k$, the distance from the Rx to the object point $k$, distance from the Tx to the $l_{t h}$ RIS element, and the distance from the the $l_{t h}$ RIS element to the object point $k$.  

We first derive the received signal corresponding to the RIS path. 
Without loss of generality, we begin with focusing on the signal propagation from the transmitter to the receiver via one particular reflecting element of the RIS, denoted by $l$, with $l \in \{ 1,...,L \}$. The phase shift induced by the $l_{th}$ RIS element is denoted as $\theta_l$. As shown in Fig.~\ref{system}, the adjustable phase shift of the RIS is realized by controlling electronic PIN diodes. Theoretically, each PIN diode can be electrically adjusted into two different states, i.e., the ON and OFF states to generate a phase shift of $\pi$ in rad. Thus, the quantization level of the discrete phase depends on the number of PIN diodes. For instance, $2$ PIN diodes that are integrated to each element enable $4$ kinds of phase shift. Thus, achieving continuously phase shifting requires a large number of PIN diodes, which is impractical for RIS. Thereby, only discrete phase shift with a limited quantization level can be realized in practical systems. Here, we consider a finite number of quantization levels $C$ with the number of PIN diodes being $b$, i.e., $C=2^b$. Note that $\theta _{l}$ is periodic with respect to $2 \pi $, thus we consider them in $[0,2 \pi)$ for convenience in the sequel of this paper. Thus, the set of discrete phase shift values of each element is given by: 
\begin{equation}
\label{equ4}
F = \{0, \frac{2\pi}{C}, ...,\frac{(C-1)2\pi}{C} \}
\end{equation}

The system is based on WiFi signal which operates under IEEE 802.11 standards. Current WiFi protocols adopt OFDM (Orthogonal Frequency-Division Multiplexing) for packet transmission to avoid performance degradation caused by interference and fading. With OFDM, the total spectrum is partitioned into multiple orthogonal subcarriers, and wireless data is transmitted over the subcarriers using the same modulation and coding scheme (MCS) to mitigate frequency selective fading. Let $x(t) = e^{-2 \pi f_{s}t}$ denote the complex-valued transmitted signal in subcarrier $s$, where $f_{s} $ is the corresponding signal frequency. The reflection of RIS element $l$ can be expressed by multiplying a complex reflection coefficient $e^{j \theta_l}$ on the signal.
The signal propagation along the RIS path introduces time-delay, the path loss and the reflection of the object, where the time-delay induces a time shift version of the transmit signal, the path loss is given by multiplying a attenuation factor $\alpha_{lk} $, and the effect of object reflection is expressed by multiplying the reflection coefficient. Combining these effects and ignoring the hardware imperfections such as circuit non-linearity and phase noise without loss of generality, the received signal of the RIS path reflected by the $l_{th}$ element can expressed as
\begin{equation}
\begin{aligned}
y_{ris,l}(t) &=\sum_{k} \alpha_{lk} v(k)e^{j\theta_l} x(t- \tau _{l}(k))\\
& =\sum_{k}\alpha_{lk} v(k)x(t) e^{j\theta_l} e^{-2 \pi f_{s} \frac{d_{ris,l}(k)}{c}},
\end{aligned}
\end{equation}
where $v(k)$ is the reflection coefficient of point $k$, $\alpha_{lk}$ is the path loss, $ \tau _{l}(k)$ is the time delay induced by propagation along the RIS path, and $c$ is the speed of light.

For simplicity, we assume that there is no signal coupling in the reflection by neighbouring RIS elements, i.e., all RIS elements reflect the incident signals independently \cite{wu_intelligent_2021}. 
Besides, we only consider signals reflected by the RIS for the first time and ignore those reflected for two or more times, which are much weaker due to the path loss. As such, the received signal from all RIS elements can be modeled as a superposition of their respective reflected signals, which is given by:
\begin{equation}
\begin{aligned}
y_{ris}(t) &=\sum_{k}\sum_{l=1}^{L} \alpha_{lk} v(k)x(t) e^{j\theta_l} e^{-2 \pi f_{s} \frac{d_{ris,l}(k)}{c}}.
\end{aligned}
\end{equation}

Another component of the received signal is the propagation along the direct path. 
By jointly considering the time-delay, the path loss and the reflection of the object, the signal accounting for the direct path is given by
\begin{equation}
\begin{aligned}
y_{de}(t) &=\sum_{k} v(k) \alpha_k x(t- \tau _{l}(k))\\
& =\sum_{k} v(k) \alpha_{k} x(t) e^{-2 \pi f_{s} \frac{d_{de}(k) }{c}},
\end{aligned}
\end{equation}
where $\alpha_{k}$ is the pass loss and $ \tau _{l}(k)$ is the time delay induced by propagation along the direct path.

As the WiFi signal is narrowband, the delays of signal propagation along the direct path and the RIS path are not resolvable, namely $d_{ris}-d_{de}<c/B$, where $B$ is the bandwidth. Hence, the received signal is the summation of signal along the direct path and the RIS path that is given by
\begin{equation}
\begin{aligned}
\label{time-domain}
y(t) &=y_{ris}(t)+y_{de}(t),\\
& =\sum_{k}v(k)x(t) {(\alpha_k e^{j\beta_{k}^{s}} + \sum_{l=1}^{L} \alpha_{lk} e^{j(\phi_{lk}^{s}+\theta_{l})})},
\end{aligned}
\end{equation}
where $\beta_{lk}^{s}$ and $\phi_{k}^{s}$ are the phase shift induced by time-delay of the direct path and RIS path, which can be expressed as
\begin{equation}
\label{equ2}
\beta_{k}^{s} = \frac{2\pi}{\lambda_{s}} (d_{tk}+d_{rk}),
\end{equation}
\begin{equation}
\label{equ3}
\phi_{lk}^{s} = \frac{2\pi}{\lambda_{s}} (d_{tl}+d_{lk}+d_{rk}),
\end{equation}
where $\lambda_{s}$ is the wavelength of the $s_{th}$ subcarrier.

In WiFi devices, rather than directly use the received signal, we could capture the Channel State Information (CSI), which characterizes the signal propagation as
\begin{equation}
\label{equ1}
h_s =\sum_{k}{v(k)(\alpha_k e^{j\beta_{k}^{s}} + \sum_{l=1}^{L} \alpha_{lk} e^{j(\phi_{lk}^{s}+\theta_{l})})}.
\end{equation}

\begin{figure}[htbp]
	\centering
	\includegraphics[width=3in]{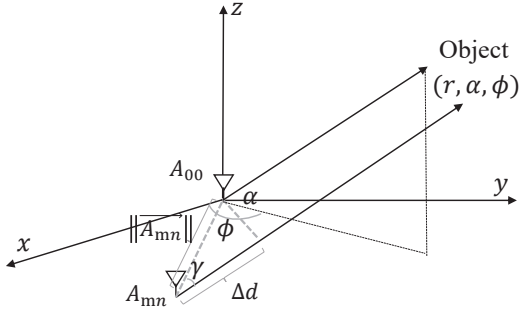}
	\caption{A geometric representation of the phase difference between antenna $A_{mn}$ and $A_{00}$. For signal reflected by the object at $(r,\alpha, \phi)$, the difference of the phase shift is related to the propagation distance difference $\Delta d$. }
	\label{delta_d}
\end{figure}

\section{RIS Beamforming}
The first step towards high-resolution imaging is to separate the signals from different spatial locations. To this end, we propose a RIS beamforming framework by controlling the phase shift of RIS elements. 
Specifically, the signals arrived at different RIS element experience different propagation distance, where the difference is denoted by $\Delta d$ in Fig.~\ref{delta_d}.
We compute the phase difference between the antenna-pair, $A_{mn}$ and $A_{00}$. The phase change in a complex wave as it traverses a distance $\Delta d$ is given 
by $e^{-j \frac{2\pi}{\lambda}\Delta d}$, where $\lambda$ is the signal wavelength. 
Using trigonometric identities, we derive the following equations
\begin{equation}
\begin{aligned}
\label{dmn}
\Delta d_{mn}(\alpha,\phi) &= || \vec{A}_{mn} || cos \gamma, \\
&= || \vec{A}_{mn} || \frac{\vec{A}_{mn} \vec{s}(\alpha,\phi)}{|| \vec{A}_{mn} || \vec{s}(\alpha,\phi) ||||},\\
&= mdsin \alpha cos\phi+nd sin\alpha sin\phi,
\end{aligned}
\end{equation}
where $\vec{A}_{mn}$ is the vector from the origin to the antenna element,
$\vec{s}(\alpha,\phi)$ is the directional vector corresponding to the signal receiving from direction $(\alpha,\phi)$, and $\gamma$ is the angle between the two vectors.

By adjusting the phase shift between all antenna elements, a 2-D antenna array can compute the signal intensity $p$ of signals arriving along the azimuth direction $\phi$ and elevation direction $\alpha$ by adding the CSI from different antennas coherently, which can be expressed as 
\begin{equation}
\label{equ5}
p(\alpha,\phi) = \sum_{m=0}^{M-1} \sum_{n=0}^{N-1} h(m,n)e^{-j \frac{2\pi}{\lambda} (mdsin \alpha cos\phi+ndsin\alpha sin\phi)},
\end{equation}
where $h(m,n)$ is the CSI, $d$ is the separation between two adjacent antennas. 

Similarly, the CSI among different subcarriers can be coherently combined according to the phase shift that is related to the range of the object. Mathematically, a set of signals of multiple subcarriers can be used to compute the signal intensity $p$ of depth $r$ as follows
\begin{equation}
\label{equ6}
p(r) = \sum_{s=1}^{S} h(s)e^{-j \frac{2\pi}{\lambda_{s}}  r},
\end{equation}
where $h(s)$ is the CSI of the $s_{th}$ subcarrier, and $\lambda_{s}$ is the corresponding frequency. 
By projecting CSI of different antenna-subcarrier pair $h(s,m,n)$
on each voxel, we can obtain the intensity of 3D voxels to form the imaging result.

In the RIS-aided WiFi imaging system, we can utilize the RIS to beamform signals in a specified voxel to generate its intensity. 
This is achieved by combining elements of the RIS and direct path in such a way that signals reflected from a particular voxel experience constructive interference while others experience destructive interference. The best performance of beamforming at point $k'$ is achieved when the RIS phases are chosen as $\theta_{l}=\phi_{lk'}^{s} = \beta_{k'}^{s} - \phi_{lk'}^{s} $, so that all the signal terms are phase-aligned. The coordinates of the Tx and point $k'$ are $(r_t,\alpha_t,\phi_t)$ and $(r_k,\alpha_k',\phi_k')$, respectively. According to (6) and (7), the chosen $\phi_{l}$ can be computed  using trigonometric identities :
\begin{equation}
\label{equ7}
\begin{aligned}
\phi_{lk'}^{s} &= \frac{2 \pi}{\lambda_s} (d_{tk'}-d_{tl}-d_{lk'}),\\
& = \frac{2 \pi}{\lambda_s}(mdsin \alpha_t cos\phi+t+ndsin\alpha_t sin\phi_t\\
&+ mdsin \alpha_k' cos\phi_k'+ndsin\alpha_k' sin\phi_k'),
\end{aligned}
\end{equation}
where $(m,n)$ is the 2D index of the $l_{th}$ RIS element. The phase-shift element in (12) has the same form with that in (9), demonstrating the superposition of all RIS elements with the chosen phase-shift can achieve direction beamforming.
     
Besides, CSI of all subcarriers with different frequency should also be summed to achieve range match-filtering. Then, the beamforming result of point $k'$ is: 
\begin{equation}
\begin{aligned}
\label{equ8}
p(k') = \sum_{s=1}^{S} (\sum_{k}{v(k)(\alpha e^{j\beta_{k}^{s}} + \sum_{l=1}^{L} \alpha_{lk} e^{j(\phi_{lk}^{s}+\phi_{lk'}^s)})}).
\end{aligned}
\end{equation}

The reflection coefficient of points in the area of interest is denoted as $\boldsymbol{v} = [v_1,v_2,...,v_K]^{T}$, and the corresponding beamforming result is denoted as $\boldsymbol{p} = [p_1,p_2,...,p_K]^{T}$. The relationship between them can be denoted as:
\begin{equation}
\label{equ9}
\boldsymbol{p} = \boldsymbol{H} \boldsymbol{v},
\end{equation} 
where $\boldsymbol{H}\in \mathbb{C}^{K \times K}$ whose element can be obtained from (13).
\begin{figure*}
	\centering
	\subfloat[Ground Truth]{
		\begin{minipage}[b]{0.23\textwidth}
			\includegraphics[width=1\textwidth]{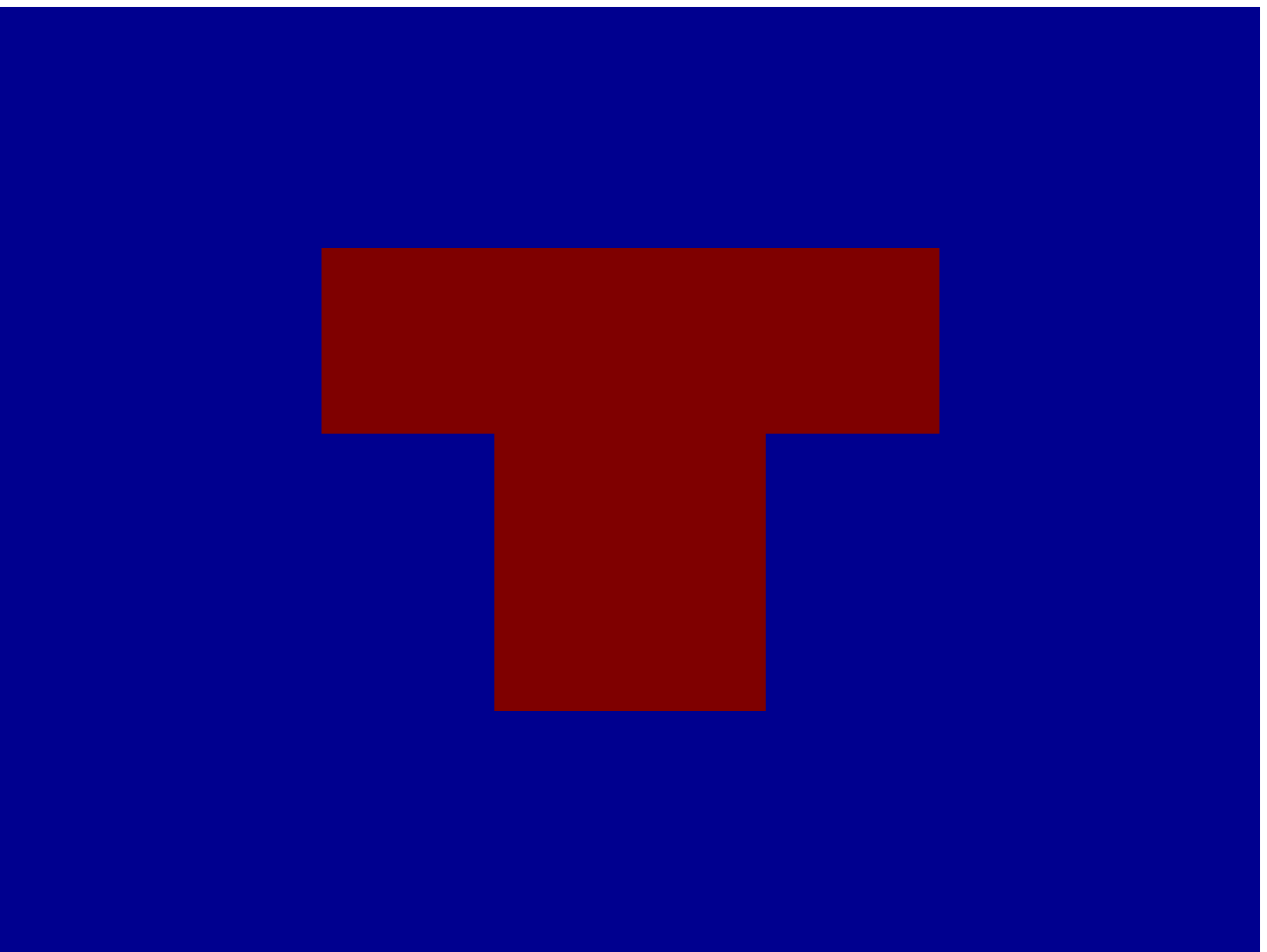} \\
			\\
			\includegraphics[width=1\textwidth]{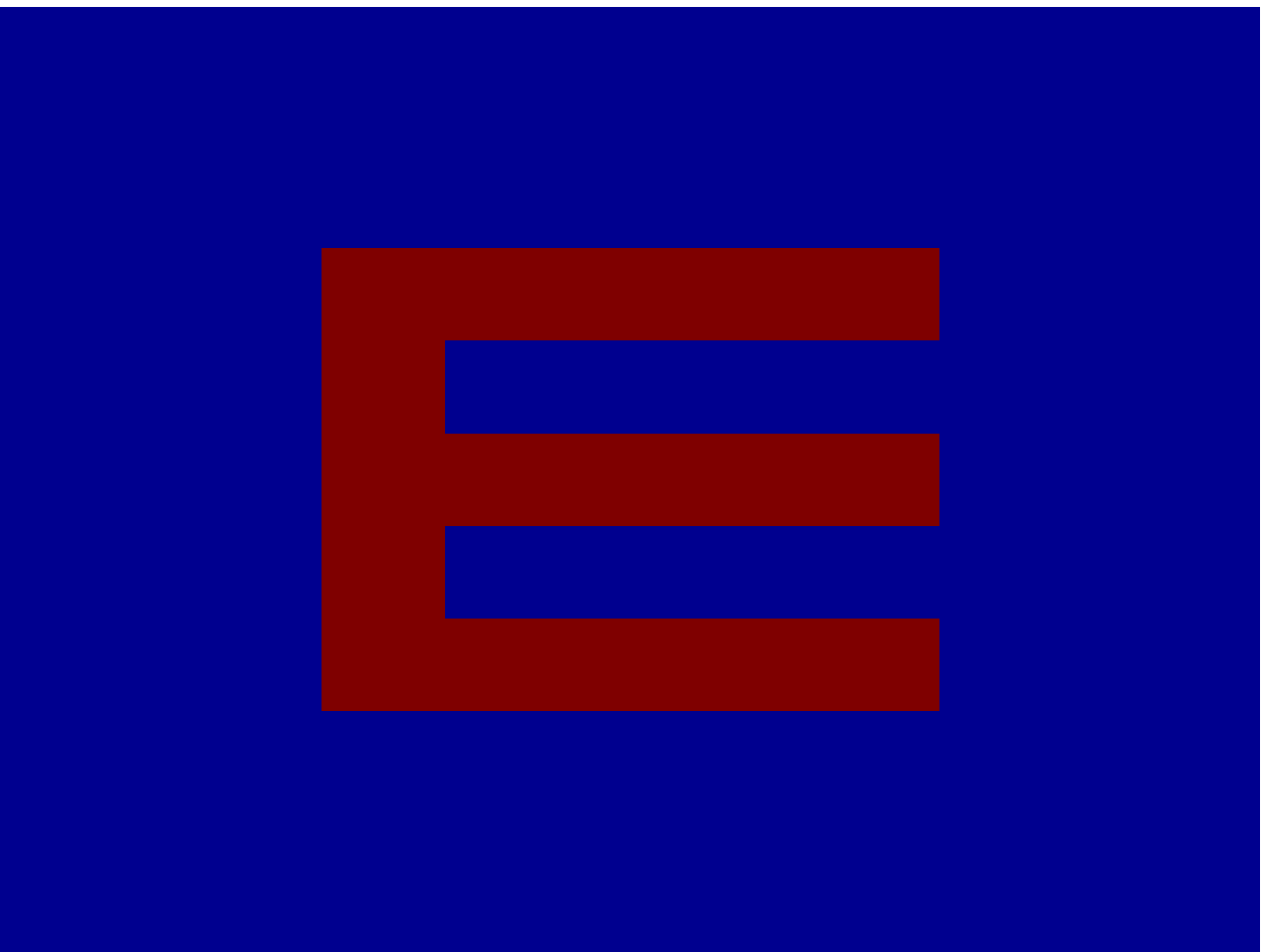}
		\end{minipage}
		\label{all_gt}
	}	
	\subfloat[Commodity WiFi]{
		\begin{minipage}[b]{0.23\textwidth}
			\includegraphics[width=1\textwidth]{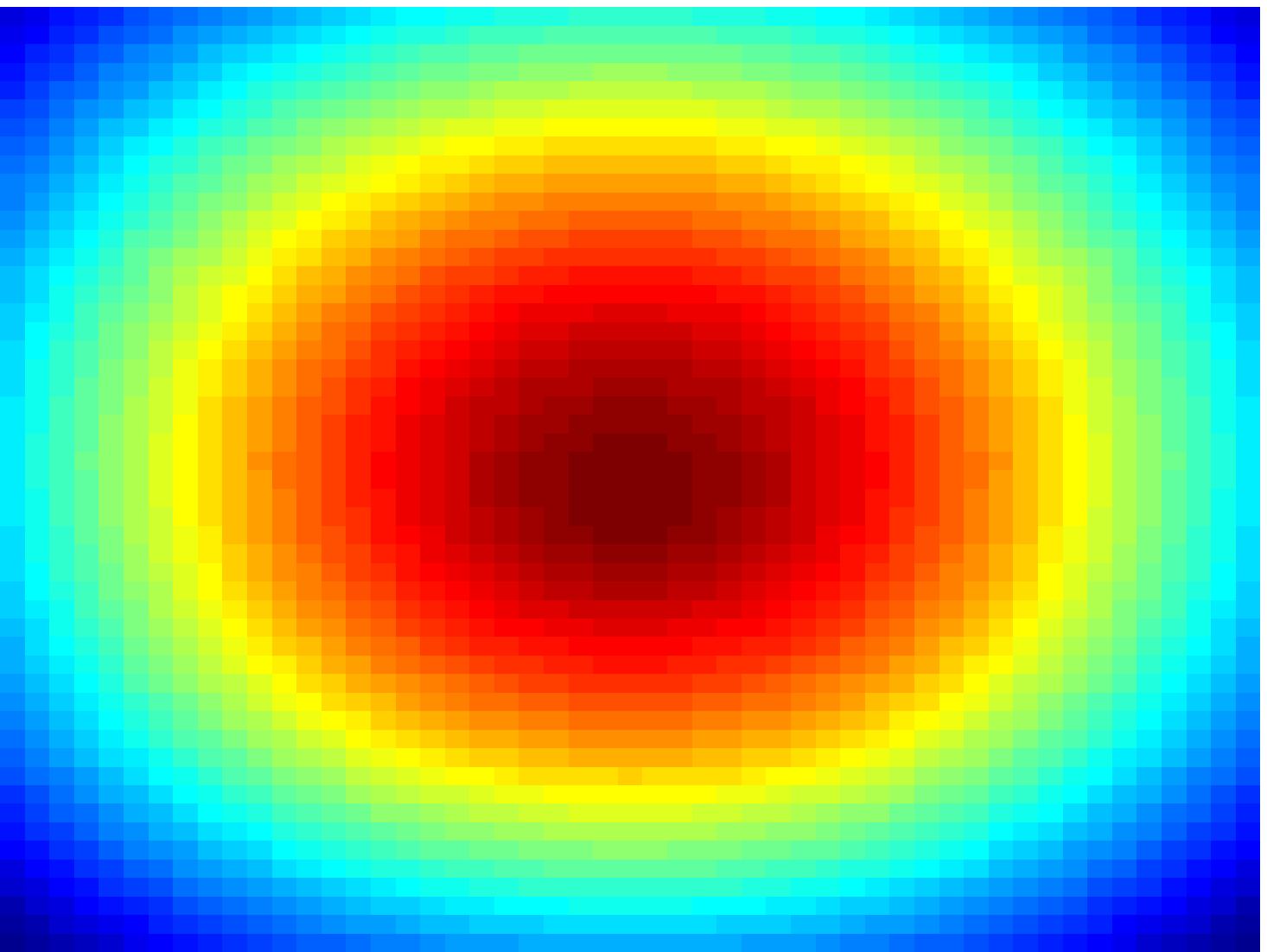} \\
			\\
			\includegraphics[width=1\textwidth]{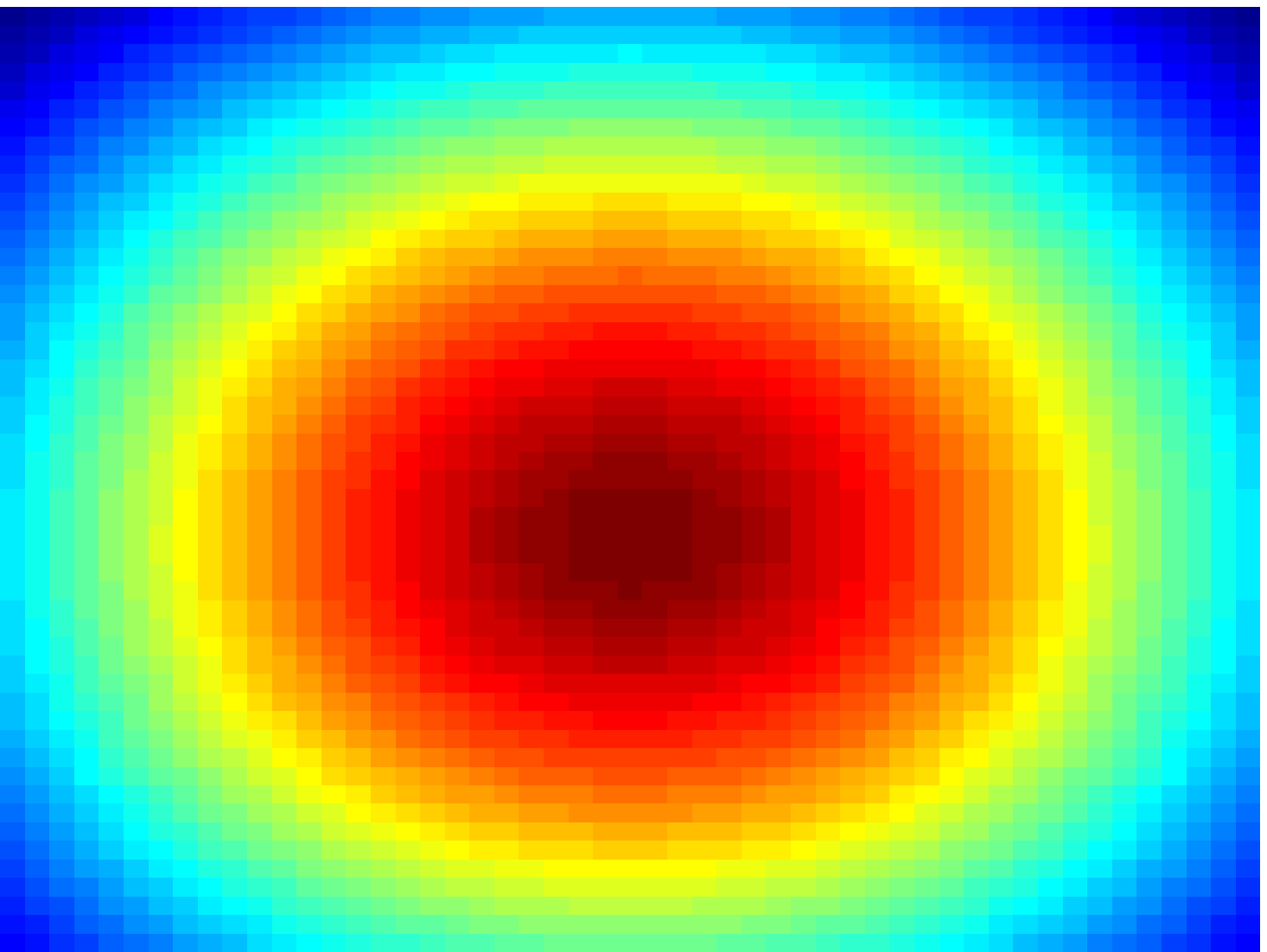}
		\end{minipage}
		\label{all_WiFi}
	}
	\subfloat[RIS-Beamforming]{
		\begin{minipage}[b]{0.23\textwidth}
			\includegraphics[width=1\textwidth]{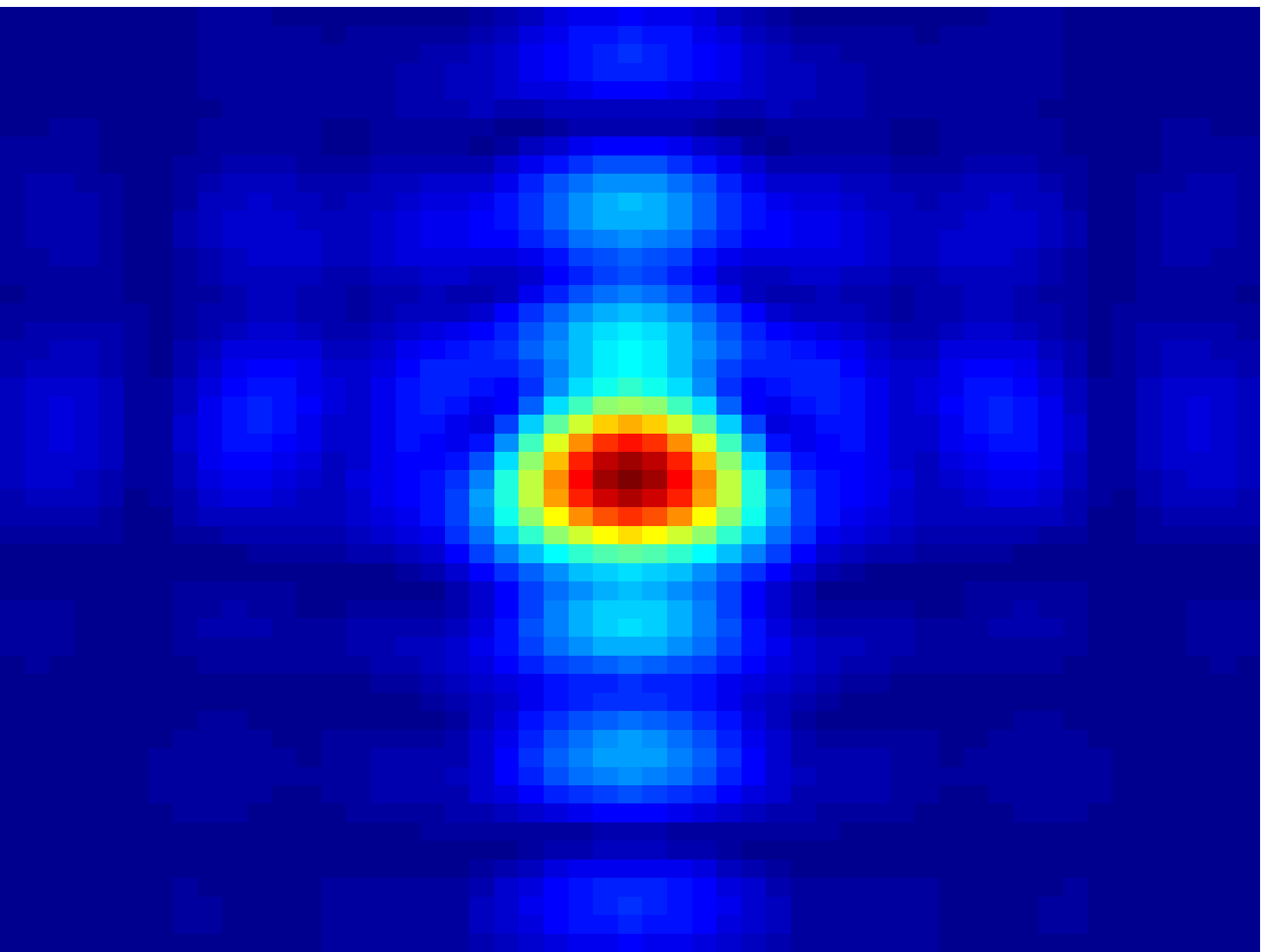} \\
			\\
			\includegraphics[width=1\textwidth]{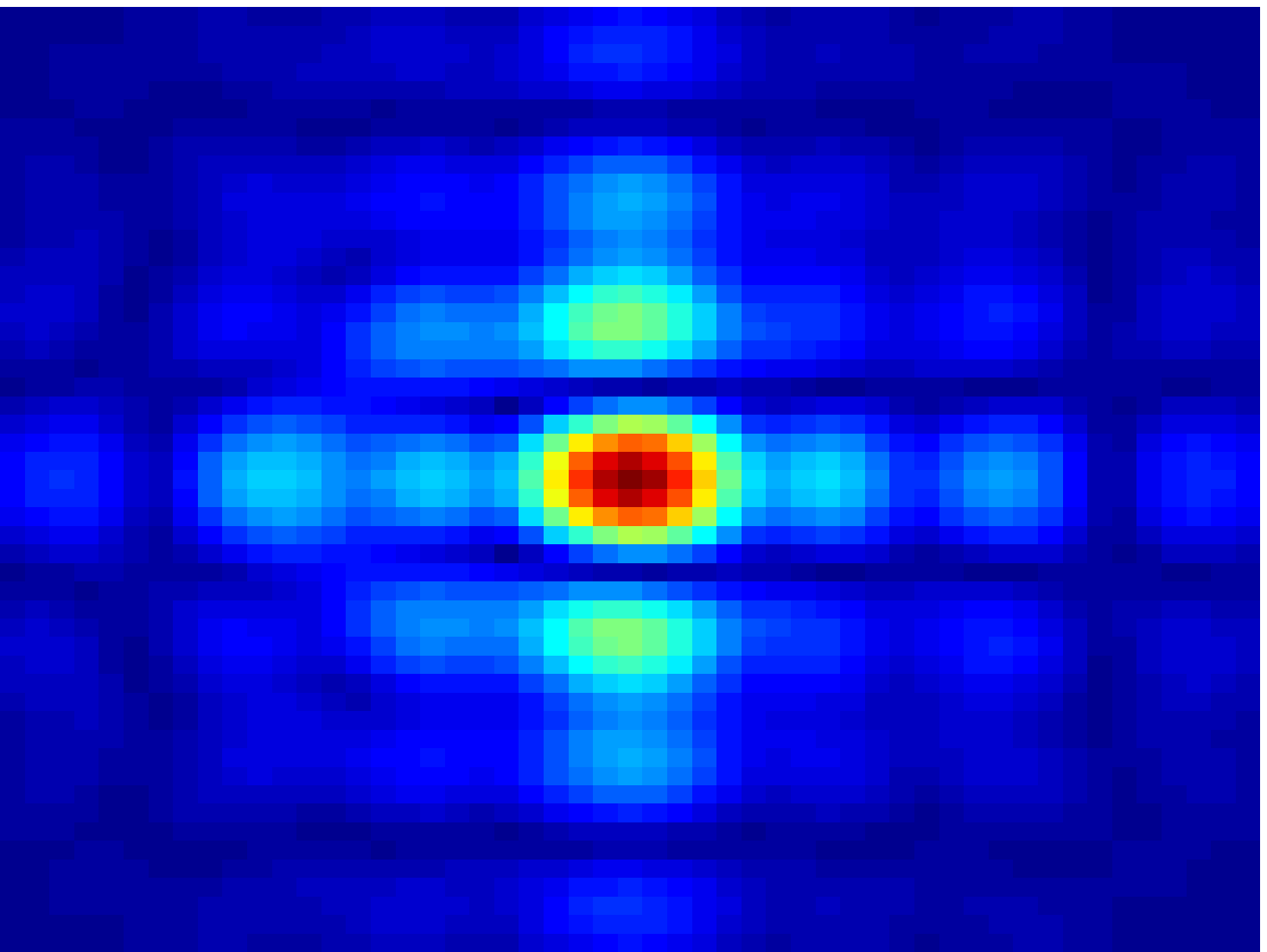}
		\end{minipage}
		\label{all_bf}
	}
	\subfloat[RIS-Optimization]{
		\begin{minipage}[b]{0.23\textwidth}
			\includegraphics[width=1\textwidth]{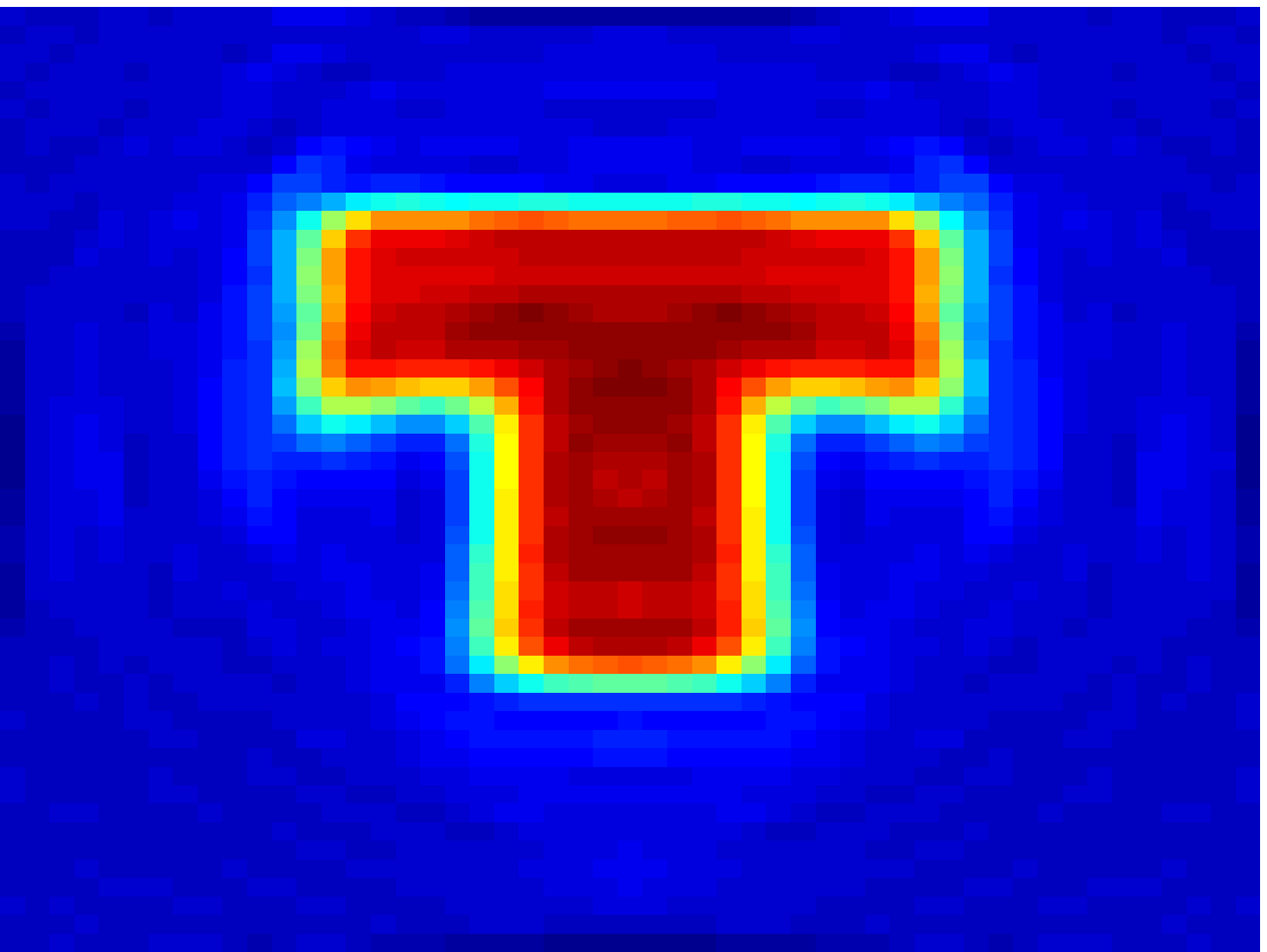} \\
			\\
			\includegraphics[width=1\textwidth]{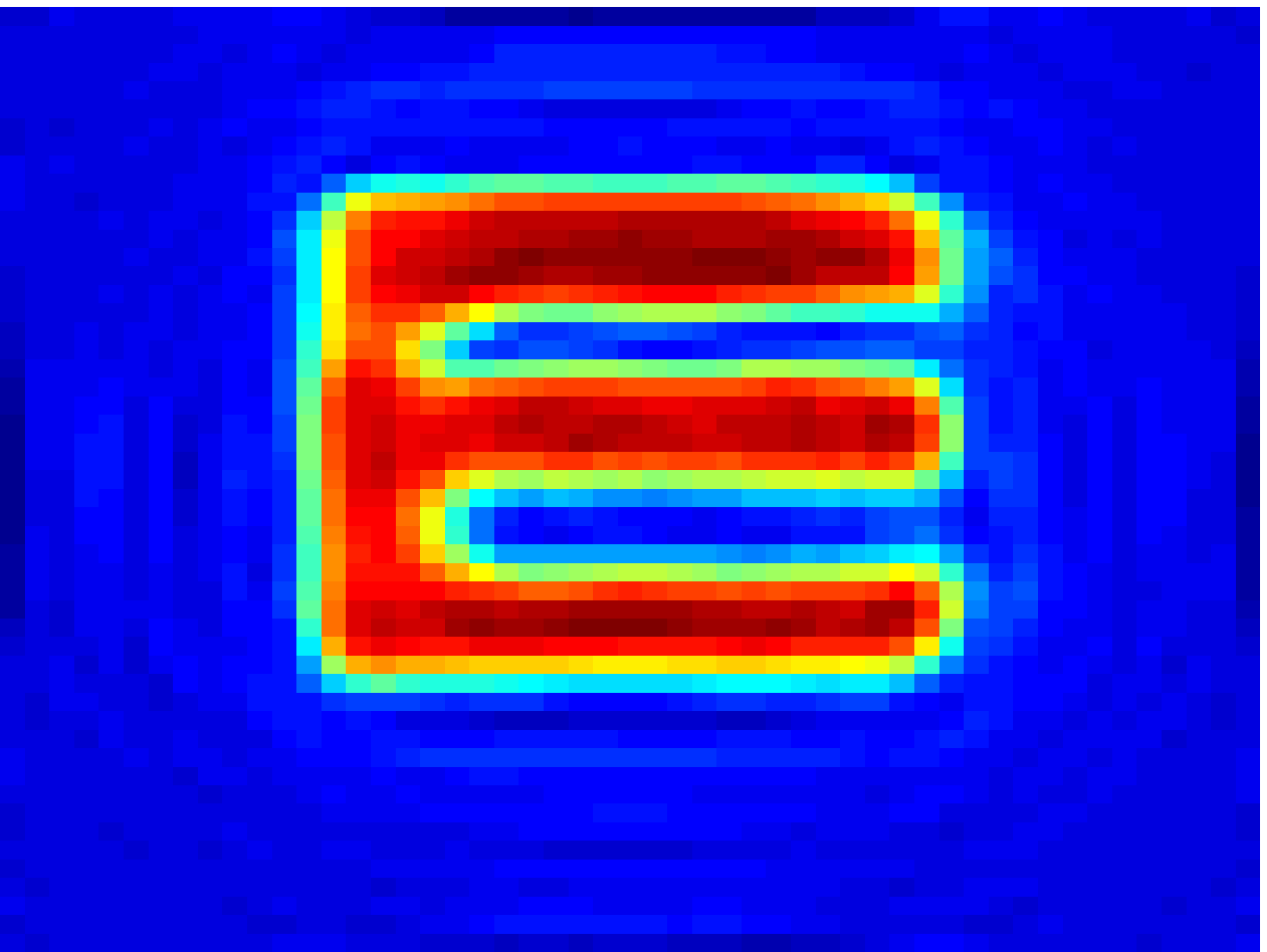}
		\end{minipage}
		\label{all_op}
	}
	\label{overall}
	\caption{The imaging results with different methods: (a) the ground truth; (b) the imaging result obtained by commodity WiFi devices with a $3\times3$ MIMO array; (c) the imaging result of the beamforming (i.e.,the first-stage of the proposed framework) where an RIS of of $17\times 17$ elements is utilized to assist the imaging; (d) the imaging result of the proposed optimization-based method where an RIS of of $17\times 17$ elements is utilized to assist the imaging. The color indicates the intensity value of reconstructed point, and the higher value, the redder color. }
\end{figure*}

\section{Low-rank Optimization}
\begin{algorithm}[htbp]
	\caption{2D reconstruction} 
	\label{alg1}
	\begin{algorithmic}[1]
		\STATE Initialize $\boldsymbol{v}=0, \boldsymbol{u_1} =0,\boldsymbol{u_2} =0;$
		\FOR{k=1:K} 
		\STATE \% Update $\boldsymbol{z1}$
		\STATE  $\boldsymbol{f} =  \boldsymbol{H}\boldsymbol{v^{k-1}}+\boldsymbol{u}_1^{k-1}$
		\STATE $\boldsymbol{z_{1}}^{k}==\frac{\boldsymbol{p}+\rho \boldsymbol{f}}{1+\rho}$ 
		\STATE \% Update $\boldsymbol{z}_2$ 
		\STATE  $\boldsymbol{f} = \boldsymbol{v}^{k-1} + \boldsymbol{u}_2^{k-1}$ 
		\STATE $\boldsymbol{F}=reshape(\boldsymbol{f},[n_{x},n_{y}])$
		\STATE Initialize $\boldsymbol{Z_{2}}^{{(k)}^{(0)}}=\boldsymbol{F}$ ,$\boldsymbol{F}^{(0)}=\boldsymbol{F}_n$ 
		\FOR{j=1:J} 
		\STATE $\boldsymbol{F}^{j}=\delta(\boldsymbol{F}^{j-1}-\boldsymbol{F}) +\boldsymbol{Z_{2}}^{{(j-1)}^{(k)}}$	   
		\FOR {each patch in $\boldsymbol{F}^{j}$}
		\STATE Find its similar patch group $\boldsymbol{F}_{m}^{j}$
		\STATE Estimate weight vector $\omega$
		\STATE $[\boldsymbol{P},\Sigma,\boldsymbol{Q}]=SVD(\boldsymbol{F}_{m}^{j})$
		\STATE $\boldsymbol{Z_{2}}_{m}^{{j}^{k}}=\boldsymbol{P}S_{\omega}(\Sigma ) \boldsymbol{Q}^T$
		\ENDFOR 	    
		\STATE Aggregating $\boldsymbol{Z_{2}}_{m}^{{j}^{k}}$ to form $\boldsymbol{Z_{2}}^{{j}^{k}}$
		\ENDFOR
		\STATE $\boldsymbol{z_2}^{(k)} = reshape( \boldsymbol{Z_{2}}^{k}, [n_{x}n_{y},1])$
		
		\STATE \% Update $\boldsymbol{u}$
		\STATE $ \boldsymbol{u}^{k} = \boldsymbol{u^{k-1}} + \boldsymbol{K}\boldsymbol{v}^{k-1}-\boldsymbol{z}^{k}$
		\STATE \% Update $\boldsymbol{v}$	
		\STATE $\boldsymbol{v}^{k} - \frac{\rho}{\mu}\boldsymbol{K}^{T}(\boldsymbol{K}\boldsymbol{v}^{k-1}-(\boldsymbol{z}^{k}-\boldsymbol{u}^{k}))$
		\ENDFOR 
	\end{algorithmic}
\end{algorithm}
Beamforming computes the approximated intensity of reconstructed regions by adding the phase-shift measurements of all RIS elements and subcarriers. 
However, the imaging performance is still limited because the beam of an antenna aperture has the shape of a cone instead of a pulse, leading to blur effect in imaging results. 
Besides, the performance of beamforming is sensitive to the quantization level of phase shifting value, which however cannot be achieved intensively since high-precision phase shift requires sophisticated design and expensive hardware. 

Given (12), the reconstruction performance can be improved by estimating $\boldsymbol{v}$ that can be formulated as the following optimization problem
\begin{equation}
\hat{v} = \mathop{\arg\min}_{\boldsymbol{v}}\lVert \boldsymbol{p}-\boldsymbol{H}\boldsymbol{v}\rVert_{F}^2.
\end{equation}

The above optimization may lead to over-smooth reconstruction. 
To prevent such effect, the prior of $\boldsymbol{v}$ should be included in the reconstruction. 
Specifically, we consider the low rank prior of the matrix $\boldsymbol{v}$ to reconstruct the objects based on low-rank matrix approximation. 
Nuclear norm minimization (NNM) is one important category of low-ranking matrix approximation methods \cite{}. 
Given a matrix $\boldsymbol{Y}$ of the degraded image, NNM aims to recover the latent matrix $\boldsymbol{X}$ while minimizing the nuclear norm of $\boldsymbol{X}$. The nuclear norm of a matrix $\boldsymbol{X}$, denoted
by $\left \| \boldsymbol{X}\right \|_*$, is defined as the sum of its singular values, i.e.,$\left \| \boldsymbol{X}\right \|_* = \sum_{i}|\sigma _{i}(\boldsymbol{X})|$, where $\sigma _{i}(\boldsymbol{X})$ denotes the $i-th$ singular value. 
Cai et al. \cite{4797640} proved that the NNM-based low-rank matrix approximation problem could be easily solved by a soft-thresholding operation on the singular values of the observation matrix. The authors in \cite{6909762} considered weighted nuclear norm minimization (WNNM) algorithm, where the singular values are assigned different weights to enhance the flexibility of NNM in dealing with many practical problems. In this paper, we include the weighted nuclear norm of the albedo $\boldsymbol{v}$ as the prior to enhance imaging results. Then the optimization with the prior is denoted as
\begin{equation}
\begin{aligned}
&\mathop{\arg\min}_{\boldsymbol{v}} \frac{1}{2}\left \|\boldsymbol{p} -\boldsymbol{H}\boldsymbol{v}\right \|_{F}^{2}+\lambda \left\| \boldsymbol{v}\right \|_{\omega,*}. \\
\end{aligned}
\end{equation}
Next, we follow the general approach of the alternate direction method of multipliers method (ADMM)\cite{8186925} and split
the unknowns while enforcing consensus in the constraints
\begin{equation}
\begin{aligned}
&\text{min} \quad \frac{1}{2} \underbrace{\left \|\boldsymbol{p} -\boldsymbol{z}_1 \right \|_{F}^{2}}_{g_{1}(\boldsymbol{z}_1)}+\underbrace{\lambda \left\| \boldsymbol{z}_2\right \|_{\omega,*}}_{g_{2}(\boldsymbol{z}_2)}\\
& \text{subject to} \underbrace{\begin{bmatrix}\boldsymbol{H} \\I\end{bmatrix}}_{K} \boldsymbol {v} - \underbrace{\begin{bmatrix}\boldsymbol{z}_1 \\\boldsymbol{z}_2\end{bmatrix}}_{\boldsymbol{z}} = 0. 
\end{aligned}
\end{equation}
The augmented Lagrangian of this objective function can be written as follows
\begin{equation}
\begin{aligned}
\mathcal{L}(\boldsymbol{v},\boldsymbol{z},\boldsymbol{y}) = & \sum_{i=1}^{2} g_{i}(\boldsymbol{z}_i),\\ &+\boldsymbol{y}^T(\boldsymbol{K}\boldsymbol{v} -\boldsymbol{z} )+\frac{\rho}{2} \left \|\boldsymbol{K}\boldsymbol{v}-\boldsymbol{z}\right \|_{F}^{2},
\end{aligned}
\end{equation}
where $\boldsymbol{y}$ is the Lagrange multiplier, and $\rho$ is the penalty factor.
For convenience, the scaled form of the augmented Lagrangian is used for ADMM by adopting scaled dual variable $u=(1/\rho)y$ instead of the Lagrange multiplier $\boldsymbol{y}$, which can be expressed as
\begin{equation}
\mathcal{L}(\boldsymbol{v},\boldsymbol{z},\boldsymbol{u})=\sum_{i=1}^{2} g_{i}(\boldsymbol{z}_i)
+\frac{\rho}{2}\left \|\boldsymbol{K}\boldsymbol{v}-\boldsymbol{z}+\boldsymbol{u}\right \|_{F}^{2}.
\label{equ10}
\end{equation}
ADMM now minimizes $ \boldsymbol{v},\boldsymbol{z},\boldsymbol{u} $ w.r.t. one variable at a time while fixing the remaining variables. The minimization is then achieved iteratively by alternately updating $\boldsymbol{v},\boldsymbol{z}_i,\boldsymbol{u}$. The key steps of this algorithm are as follows:\\
Update $\boldsymbol{z}_1$, and  $\boldsymbol{f} =  \boldsymbol{H}\boldsymbol{v}+\boldsymbol{u}_1$.
\begin{equation}
\begin{aligned}
\boldsymbol{z}_1&=\mathop{\arg\min}_{\boldsymbol{z}_1} g_{1}(\boldsymbol{z}_1)+\frac{\rho}{2}\left \|\boldsymbol{f}-\boldsymbol{z}_1\right \|_{F}^{2}\\
&=\frac{\boldsymbol{p}+\rho \boldsymbol{f}}{1+\rho}
\end{aligned}
\end{equation}
Update $\boldsymbol{z}_2$, and $\boldsymbol{f}=\boldsymbol{v}+\boldsymbol{u}_2$.
\begin{equation}
\begin{aligned}
\boldsymbol{z}_2 &=\mathop{\arg\min}_{\boldsymbol{z}_2}g_{2}(\boldsymbol{z}_2)+\frac{\rho}{2}\left \|\boldsymbol{f}-\boldsymbol{z}_2\right \|_{F}^{2}\\
&=\boldsymbol{P}S_{\omega}(\Sigma ) \boldsymbol{Q}^T
\end{aligned}
\end{equation}
where $\boldsymbol{f }$ is reshaped to 2-D matrix $\boldsymbol{F}$ to compute its SVD,  $\boldsymbol{F}=\boldsymbol{P}\Sigma\boldsymbol{Q}^T$, and $S_{\omega}(\Sigma)_{ii}=max(\Sigma_{ii}-\omega_i,0)$ is a threshold function. \\
Update $\boldsymbol{u}$
\begin{equation}
\boldsymbol{u} = \boldsymbol{u} + \boldsymbol{K}\boldsymbol{v}-\boldsymbol{z}
\end{equation} 
Update $\boldsymbol{v}$
\begin{equation}
\begin{aligned}
\boldsymbol{v} &=\underset {\boldsymbol{v}} {\text{arg min}}\frac{\rho}{2}\left \|\boldsymbol{K}\boldsymbol{v}-\boldsymbol{z}+\boldsymbol{u}\right \|_{F}^{2}\\
&
= \boldsymbol{v} - \frac{\rho}{\mu}\boldsymbol{K}^{T}(\boldsymbol{K}\boldsymbol{v}-(\boldsymbol{z}-\boldsymbol{u})) 
\end{aligned}
\end{equation} 
where $\mu$ controls the step size of gradient descent. 

Note that both $g_1$ and $g_2$ are convex, ADMM is guaranteed to find the global minimum of the objective. The implementation of the reconstruction algorithm is summarized in  Algorithm \ref{alg1}.

\section{Simulation Results}

In this section, we conduct simulations to evaluate the performance of the proposed imaging algorithms. 
\subsection{Simulation setup}
For comparison, we build a WiFi imaging system based on existing methods and a RIS-aided WiFi imaging system proposed in this paper to conduct object imaging. The result of the commodity WiFi imaging system is used to validate the performance enhancement provided by RIS. 

\begin{figure}[htbp]
    \centering
	\subfloat[]{\includegraphics[width=1.5in]{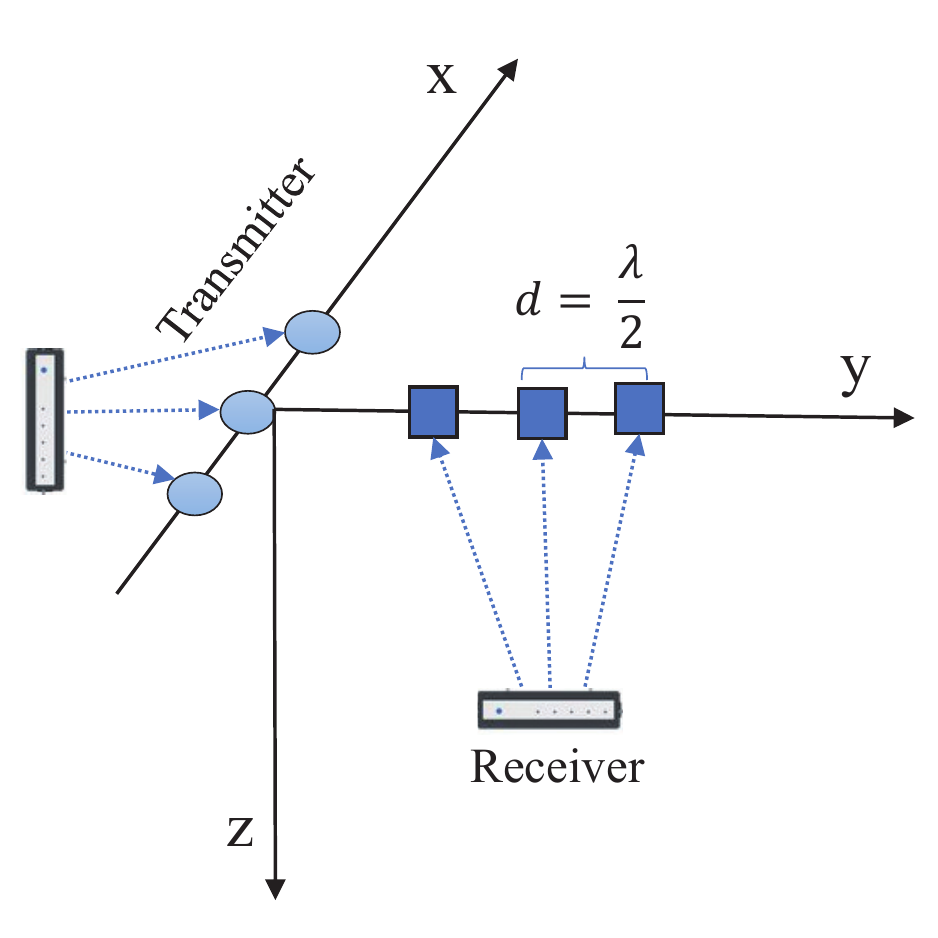}%
	\label{wifi}}
     \hfil
     \subfloat[]{\includegraphics[width=1.5in]{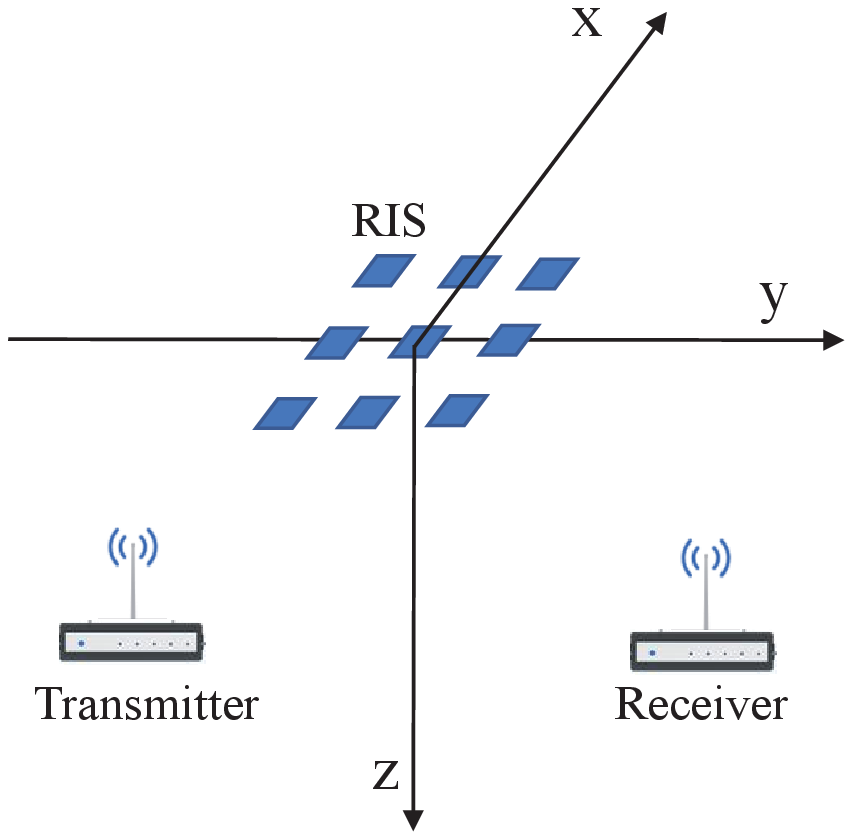}%
	\label{wifi_ris}}
    \caption{An illustration of the experiment setup: (a) shows the commodity WiFi imaging system with two APs equipped with three antennas, which form a T-shaped MIMO array for imaging; (b) shows the RIS-aided WiFi imaging system, where an RIS is equipped on the x-y plane to enhance the imaging ability of two single antenna APs.}
\end{figure}

\textbf{Commodity WiFi imaging system.} We simulate an imaging system with off-the-shelf WiFi devices as the benchmark. Usually, the off-the-shelf WiFi device is equipped with 3 omnidirectional antennas, and thus two APs can form a 3 $\times$ 3 MIMO array as shown in Fig.~\ref{wifi}. The MIMO array is a T-shaped antenna array composed of two linear arrays placed along the x-axis and the y-axis. Each linear array contains 3 antennas spacing 3cm apart. 

\textbf{RIS-aided WiFi imaging system.} RIS-aided WiFi imaging system is shown in Fig.~\ref{wifi_ris}, which contains two single-antenna APs as the transmitter and the receiver, respectively. The transmitter is located at $(0, -1, 0.5)$ and the receiver is located at $(0, 1, 0.5)$. A 2-D RIS is implemented on the x-y plane with its center at the origin.  The number of RIS elements is $M\times\ N $, where $M $ and $N $ denote the number of reflecting elements along x-axis and y-axis, respectively. 

\textbf{Signal Generation.} We simulate the WiFi signal which operates under 802.11n standard, where the center frequency is configured at 5.31GHz, and 30 subcarriers are uniformly sampled in a 40MHz band.

\subsection{Imaging Performance}
We first evaluate the imaging performance by reconstructing capital letters with an RIS of $17\times17$ elements. 
A T-shaped object and a E-shaped object are chosen and the ground truth as shown in Fig. 3, where the size is $50 \times 50$ cm, sampled as 26$\times$26 points with a step of 2 cm. Fig.~\ref{all_WiFi} plots the results obtained by commodity WiFi imaging system illustrated in Fig.~\ref{wifi}. The $3\times3$ MIMO array can only reconstruct the object as a whole with bubble-like imaging results. The results of T-shaped object and a E-shaped object are nearly the same with each other, failing to provide any detailed information. Fig.~\ref{all_bf} and Fig.~\ref{all_op} are the results obtained with the assistance of an RIS with $17\times17$ elements. Fig.~\ref{all_bf} shows the results of beamforming, while Fig.~\ref{all_op} are the results of the proposed optimization-based algorithm. It can be seen that the optimization-based algorithm with RIS achieves the best imaging results that is even comparable with the ground truth. The beamforming with RIS achieves better results than WiFi imaging since two objects can be distinguished from the imaging results, which are however far from the ground truth. These visual comparisons demonstrate the advantages of the proposed optimization-based algorithm with RIS. 

Then, we quantitatively evaluate the precision of the reconstruction using the root mean square error (RMSE) and the structural similarity (SSIM). The RMSE is defined as follows
\begin{equation}
RMSE = \sqrt{ \frac{\sum_{1}^{K}(x_{k_{rec}}-x_{k_{gt}})^2}{K}},
\end{equation}
where $x_{k_{rec}}$ and $x_{k_{gt}}$ are the $k_th$ point values of the reconstructed image and the ground truth image, $K$ is the total number of points to be reconstructed in the area of interest.
The SSIM is aimed at comparing the structural similarity between the reconstructed image and the ground truth \cite{guo2012edge}. The definition of SSIM is as follows
\begin{equation}
SSIM = \frac{(2 \mu_A \mu_B +c_1)(2 \sigma_{AB}+c_2)}{(\mu_A^2+\mu_B^2+c1)(\sigma_A^2+\sigma_B^2+c2)},
\end{equation}
where $\mu_A$ and $\mu_B$ are the average pixel values of the reconstruction image and the ground truth, respectively, $\sigma_A$ and $\sigma_B$ are the variances of the reconstruction image and the
ground truth, respectively, $\sigma_{AB}$ is the covariance between the reconstruction image and the ground truth, $c1$ and $c2$ are the constants used to stabilize the equation. Here we set $c1= (k_1 D)^2$, $c2= (k_2 D)^2$, where $D$ is the dynamic range of intensity values. The SSIM value is from 0 to 1, and if the value is closer to 1, the reconstructed image has more similar structure to the ground truth image.
\begin{figure}[htbp]
\centering
	\subfloat[RMSE]{\includegraphics[width=0.49\linewidth]{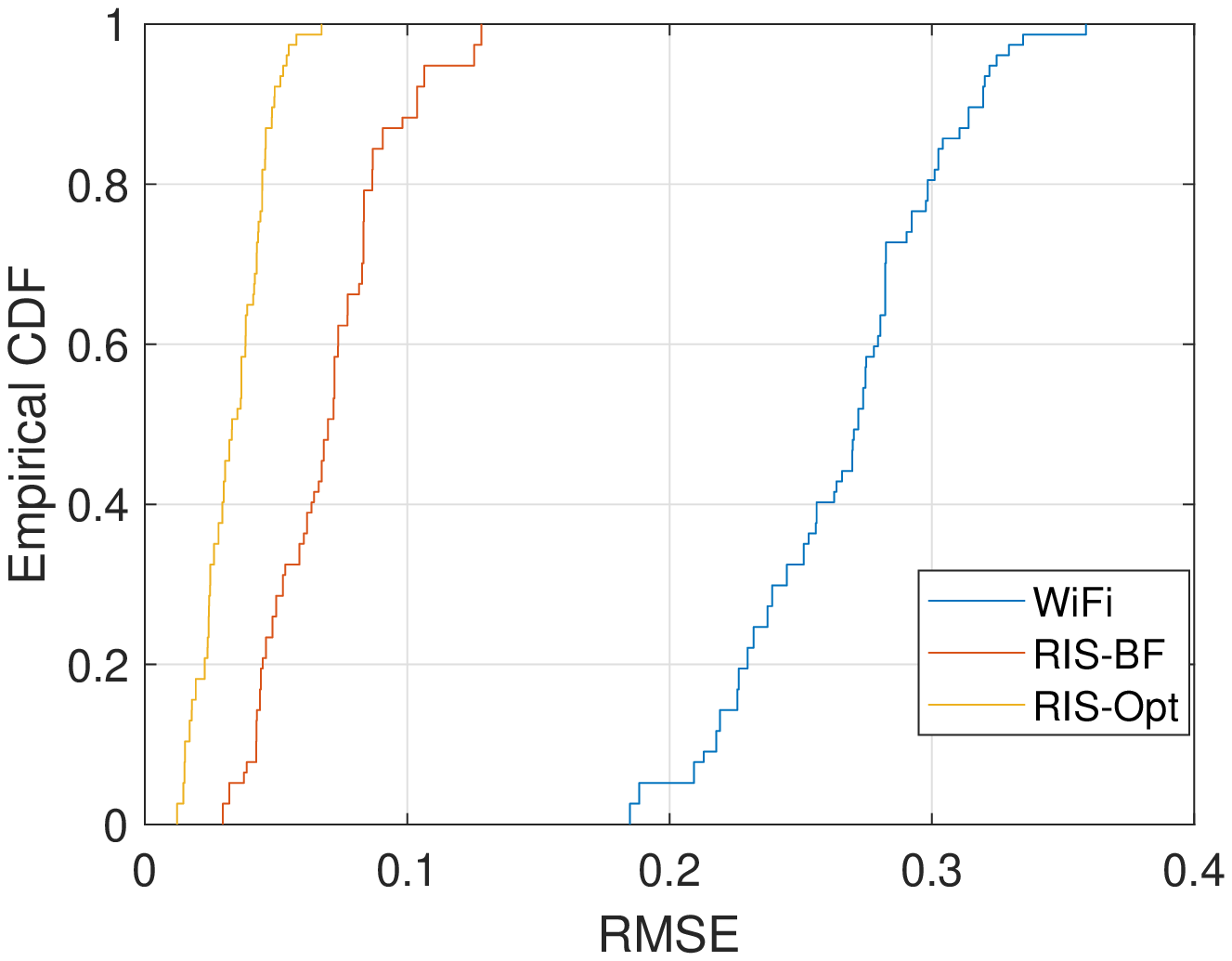}%
	\label{rmse}}
     \hfil
     \subfloat[SSIM]{\includegraphics[width=0.49\linewidth]{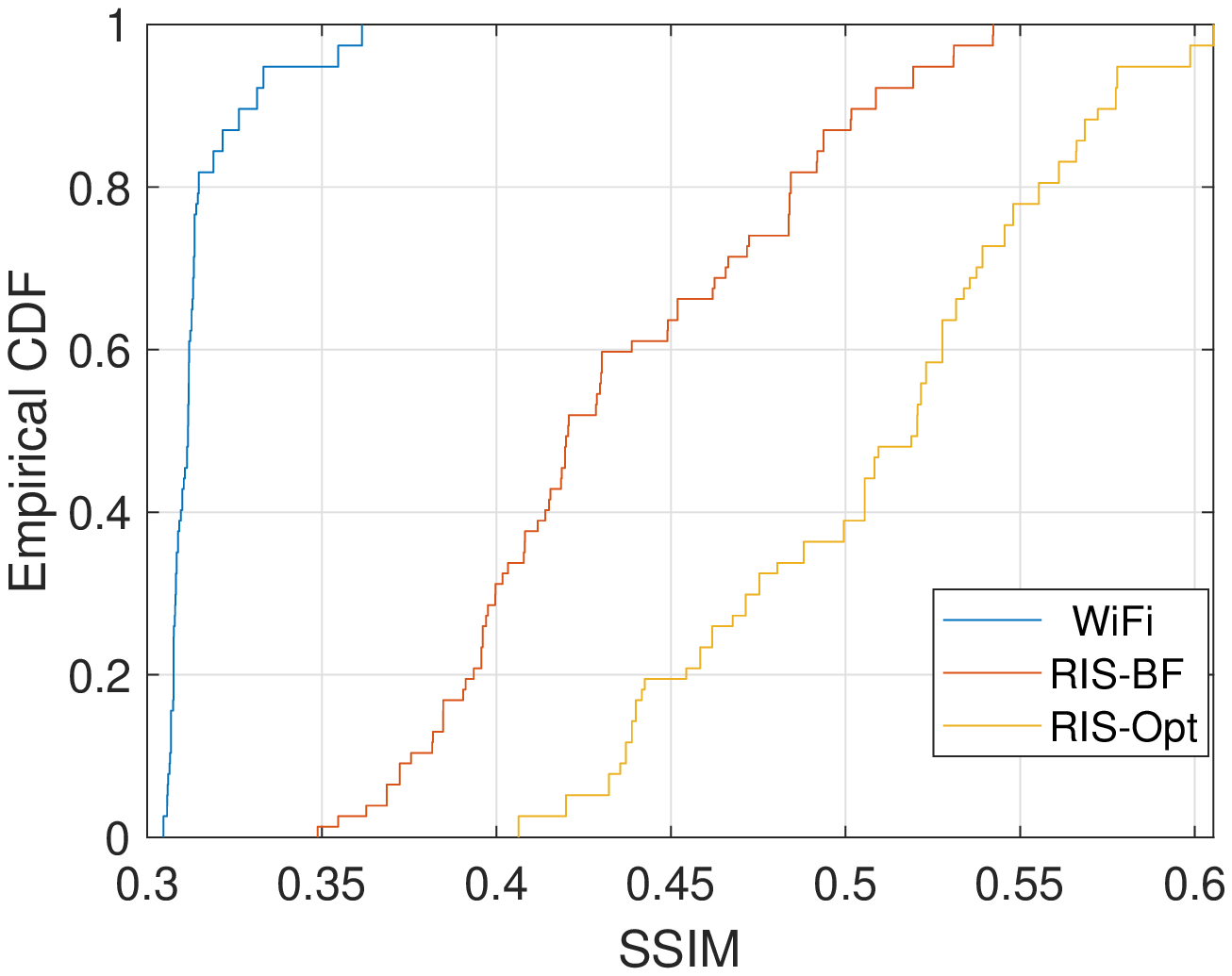}%
	\label{ssim}}
    \caption{Performance of object imaging with different systems and algorithms. }
    \label{cdf}
\end{figure}

Fig.~\ref{rmse} and Fig.~\ref{ssim} show the imaging quality of of different systems and algorithms w.r.t. RMSE and SSIM, respectively. The optimization-based algorithm (RIS-Opt) achieves the best performance with a median of 0.03 for RMSE and 0.52 for SSIM. The beamforming (RIS-BF) comes in the second place, which achieves a of 0.07 for RMSE and 0.42 for SSIM. By contrast, the median RMSE and SSIM degrade to 0.28 and 0.31 when levering commodity WiFi imaging system. 

\textbf{Impact of the number of RIS elements}
Here, we analyze the impact of the number of RIS elements. The RIS serves as a 2-D Uniform Rectangular Array (URA) in our WiFi imaging system, and the imaging performance is related to the size of RIS since the resolution is determined by the aperture size. Theoretically, the cross-range resolutions for 2-D URA are defined as
\begin{equation}
\delta_{x} = \frac{\lambda_{c}R}{D_x},  
\label{d_x} 
\end{equation}
\begin{equation}
\delta_{y} = \frac{\lambda_{c}R}{D_x},
\label{d_y}
\end{equation}
where $\lambda_{c}$ is the wavelength at the center frequency, $R$ is the distance between the surface and the antenna plane along the $z$-axis, $D_{x}$ is the antenna aperture along the $x$-axis, and  $D_{y}$ is aperture along the $x$-axis. Since the RIS acts as a URA to enable angular resolution in the WiFi imaging system, $D_{x}$ and $D_{y}$ are the length and width of the RIS, respectively.  

\begin{figure*}[htbp]
	\centering
	\subfloat[RIS: $9\times9$]{\includegraphics[width=2in]{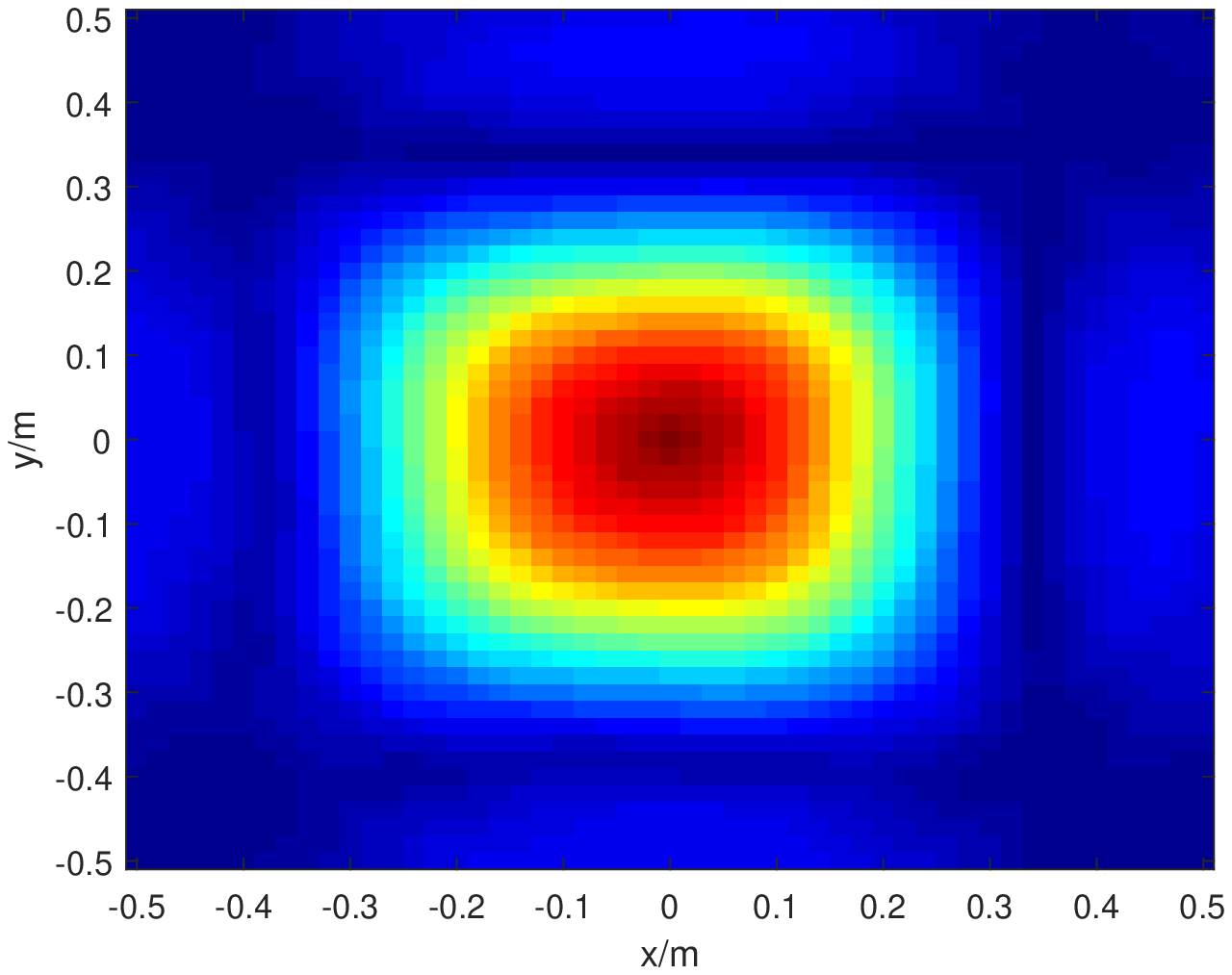}%
		\label{bf_9}}
	\hfil
	\subfloat[RIS: $11\times11$]{\includegraphics[width=2in]{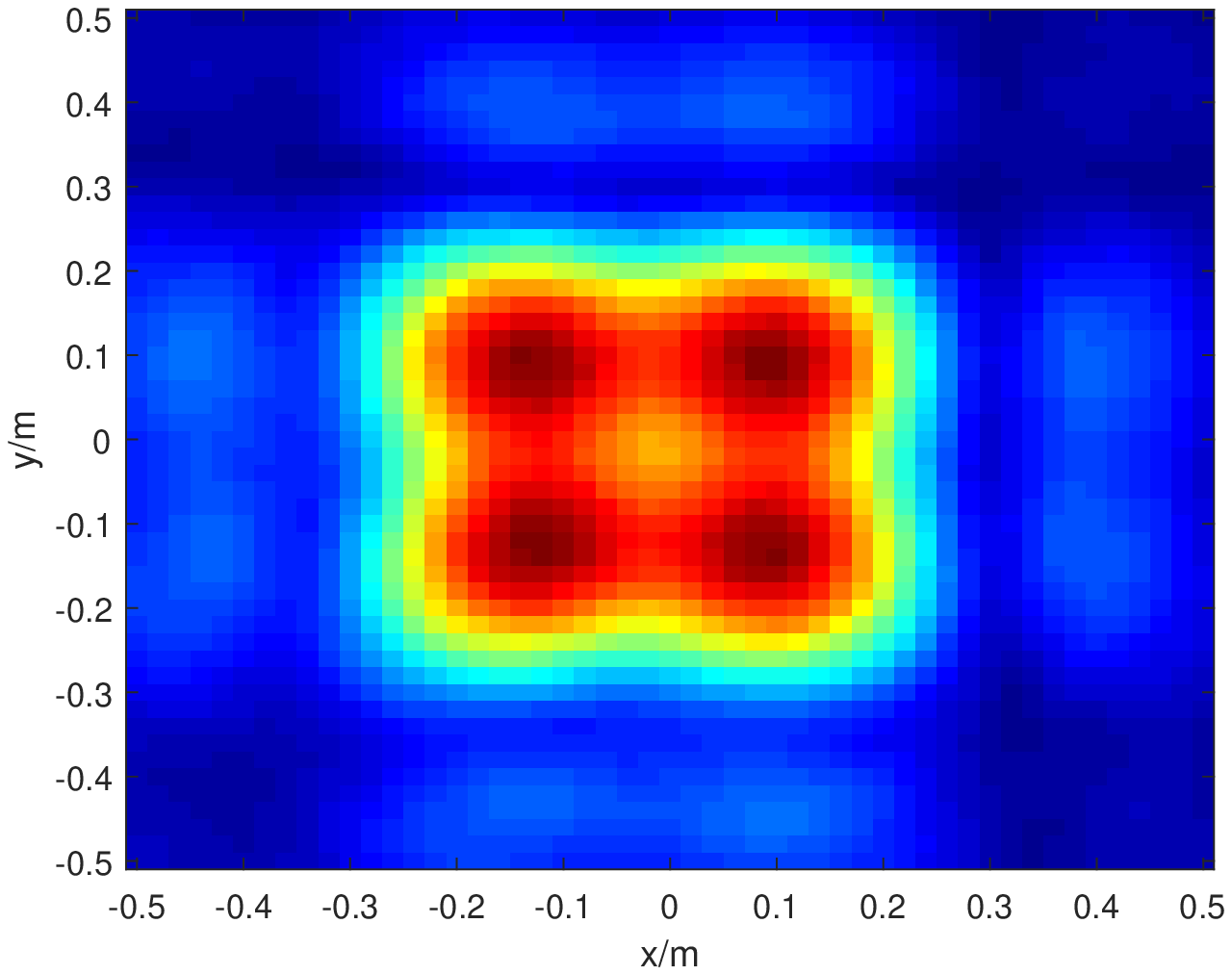}%
		\label{bf_11}}
	\hfil
	\subfloat[RIS:$13\times13$ ]{\includegraphics[width= 2in]{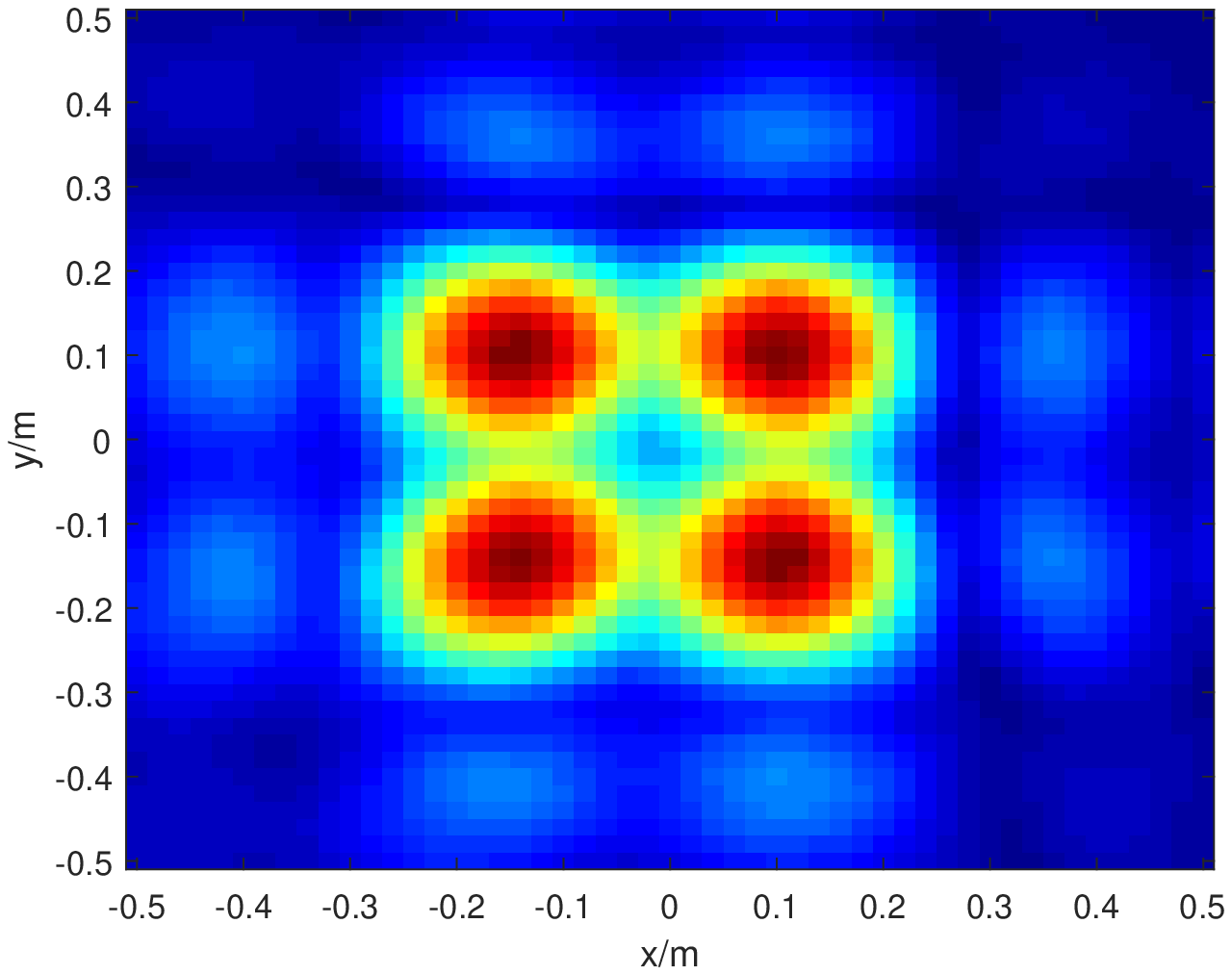}%
		\label{bf_13}}	
	\caption{The number of RIS elements versus the imaging performance of beamforming, where the ground truth is four points with spacing of 22cm.}
	\label{res_bf}
\end{figure*}

\begin{figure*}[htbp]
	\centering
	\subfloat[RIS: $9\times9$]{\includegraphics[width=2in]{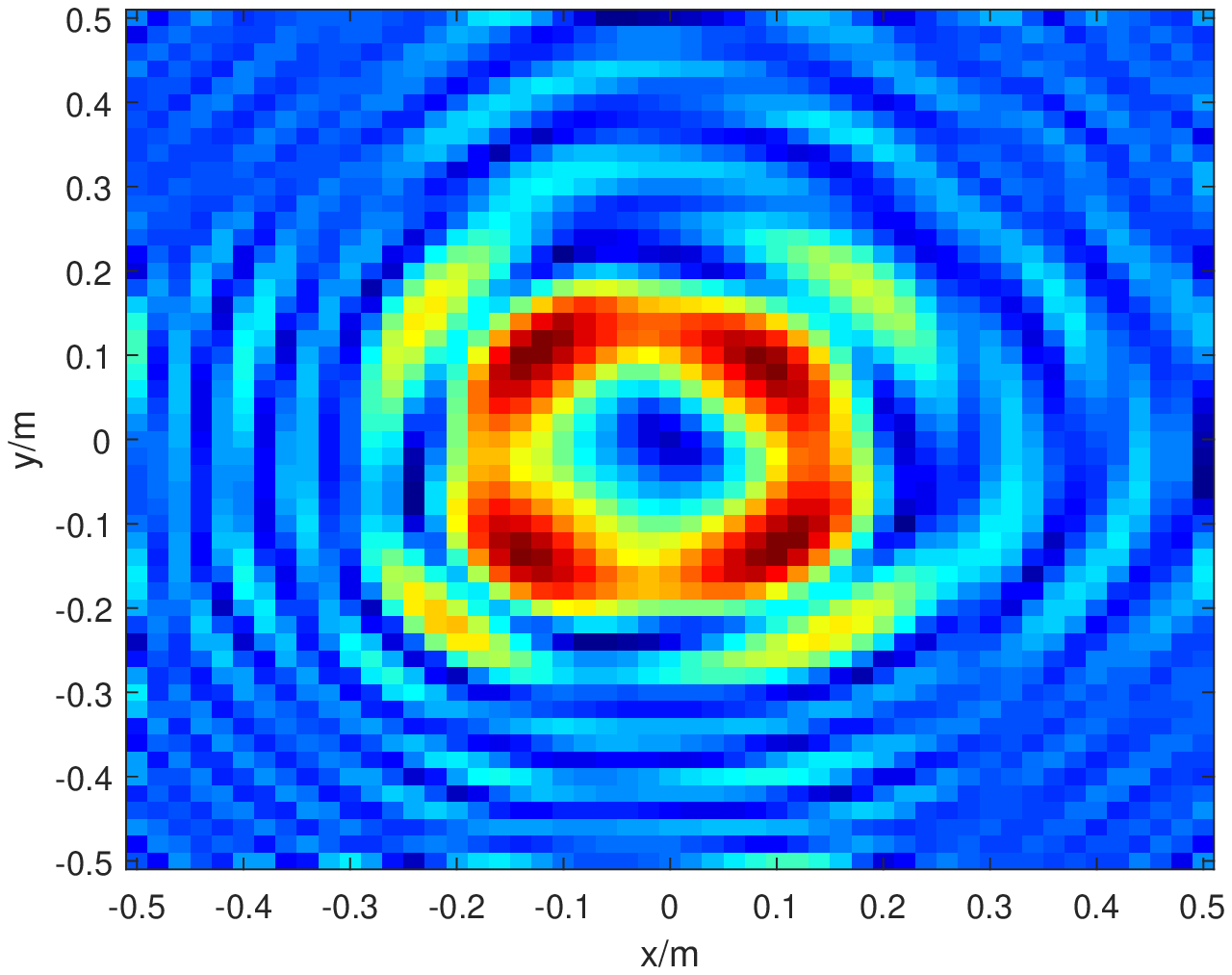}%
		\label{op_9}}
	\hfil
	\subfloat[RIS: $11\times11$]{\includegraphics[width=2in]{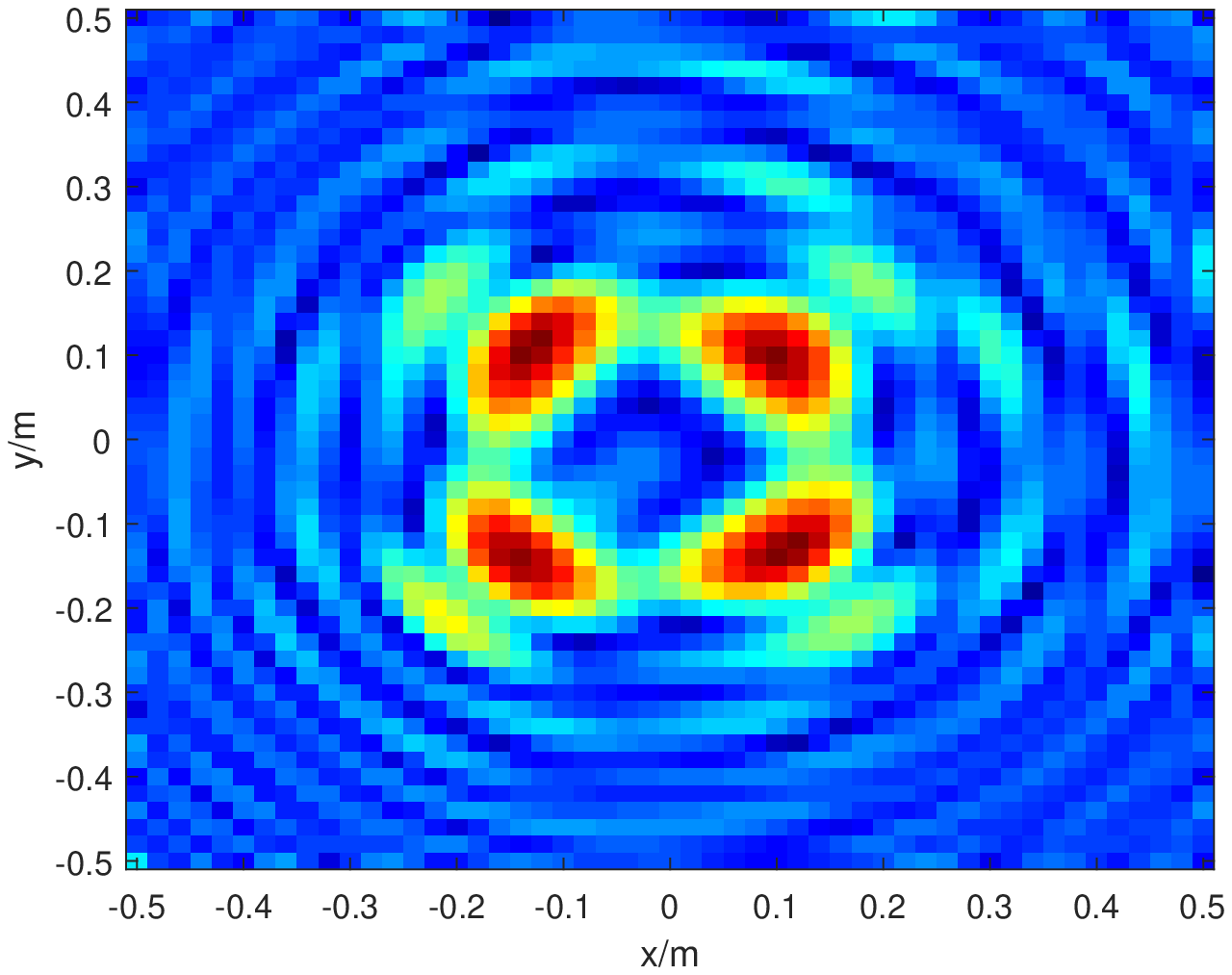}%
		\label{op_11}}
	\hfil
	\subfloat[RIS:$13\times13$ ]{\includegraphics[width= 2in]{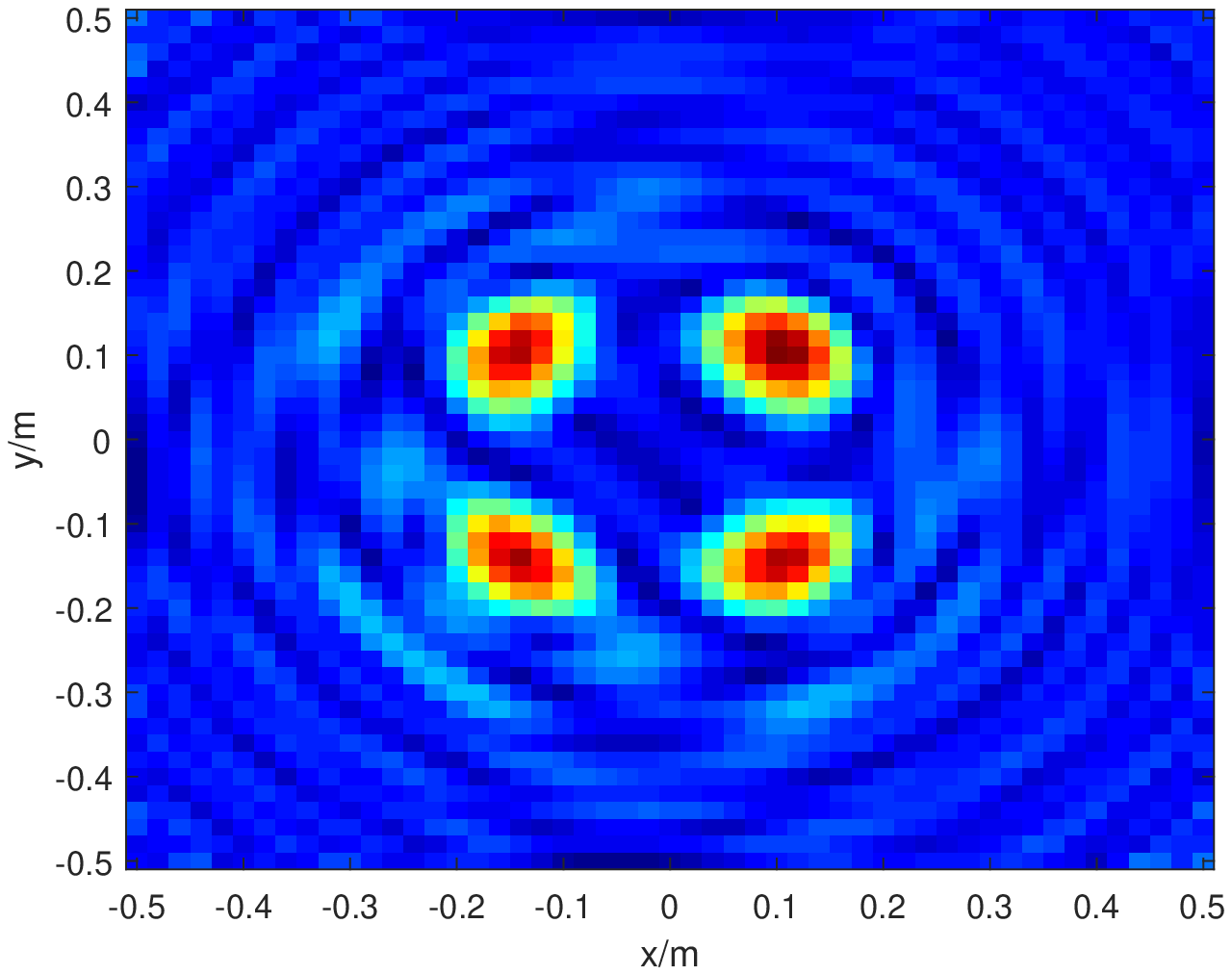}%
		\label{op_13}}	
	\caption{The number of RIS elements versus the imaging performance of optimization, where the ground truth is four points with spacing of 22cm.}
	\label{res_op}
\end{figure*}

In order to illustrate the resolution of imaging intuitively, we simulate the imaging of point-like target. We build the ground truth containing four points which form a square with side length of 22 cm.  
These points are located in a plane at $z = 1m$, which means the $R$ in (\ref{d_x}) and (\ref{d_y}) is equal to 1m. The spacing of adjacent RIS elements is half wavelength, thus according to (\ref{d_x}) and (\ref{d_y}), at least 10 elements on each side of RIS are required to distinguish these four points. We show the imaging result of beamforming and optimization with different RIS element number in Fig.~\ref{res_bf}, where Fig.~\ref{bf_9}, Fig.~\ref{bf_11} and Fig.~\ref{bf_13} are the results of beamforming obtained with $9 \times 9 $, $11 \times 11 $, and $13 \times 13 $ RIS elements, respectively. We can see that these results are consistent with the theoretic analysis where only when the RIS element number is more than  $10 \times 10 $, the four point can be distinguished. In addition, as the number of RIS elements increases, the imaging results suffer less diffusion. Fig.~\ref{op_9}, Fig.~\ref{op_11} and Fig.~\ref{op_13} plot the result of optimization obtained with $9 \times 9 $, $11 \times 11 $, and $13 \times 13 $ RIS elements, respectively. We can observe that the results of optimization are better than those of beamforming due to its super-resolution imaging capability. 

\textbf{Impact of Discrete Phase-Shift Levels}
One challenge of designing practical WiFi imaging system with RIS lies in the discrete phase-shift levels of each RIS element. In such a case, instead of continuously tuning the phase shift of the RIS elements, a practical approach is to approximate the phase-shift values to their nearest values in the discrete sets. Let $b$ denote the number of bits used as phase controller, then the number of phase shift levels can be denoted as $C = 2^b$. For practical RISs which usually have massive elements, it is more cost-effective to implement discrete phase-shift levels with a small number of control bits, for instance, $b = 1$ or $b = 2$. In the following, we will show the effect of phase-shift level by comparing the imaging result with $b = 1$, $b = 2$, and $b \to \infty $(continuous phase shift). 

\begin{figure*}[htbp]
	\centering
	\subfloat[$b=1$]{\includegraphics[width=2in]{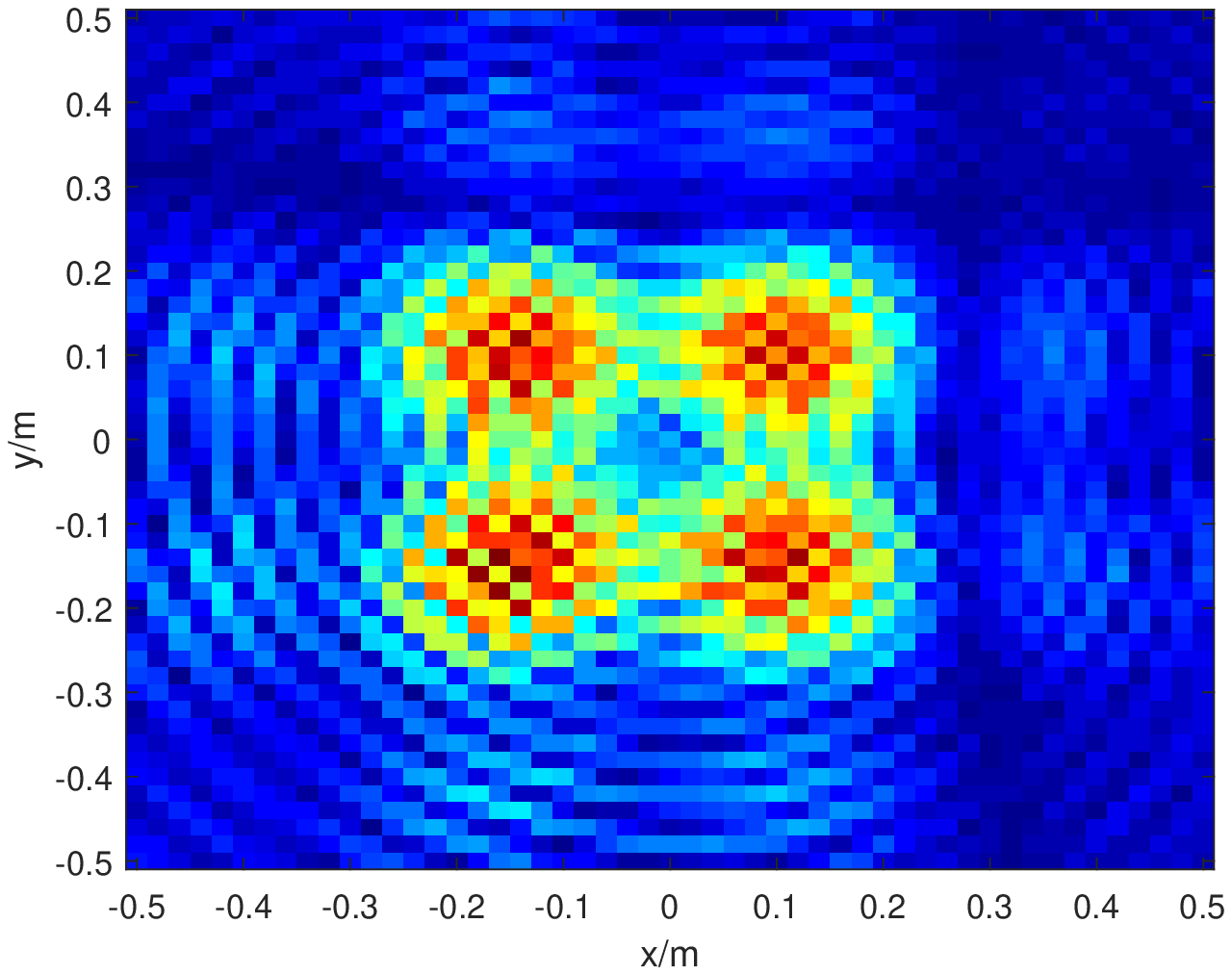}%
		\label{dis_bf_1}}
	\hfil
	\subfloat[$b=2$]{\includegraphics[width=2in]{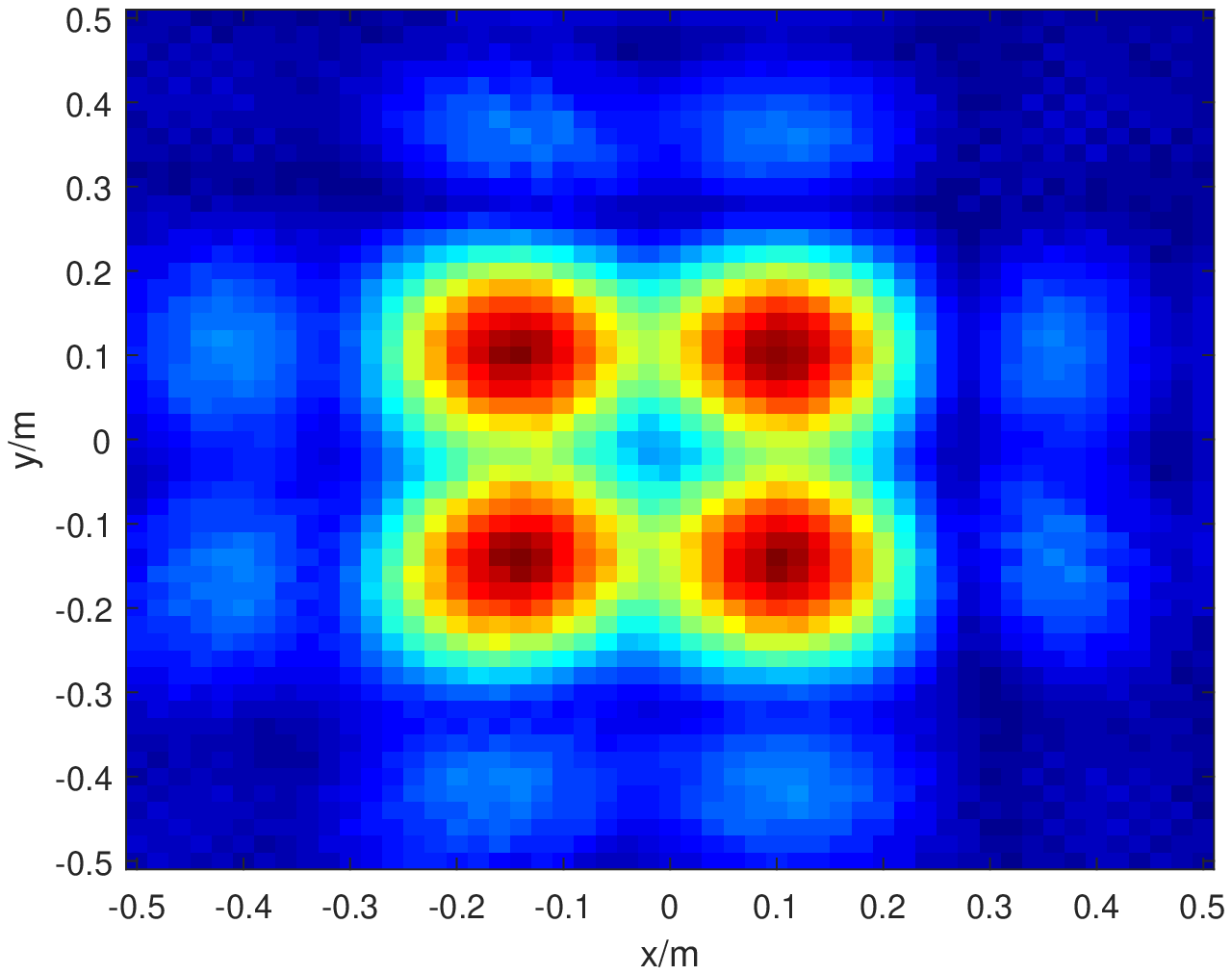}%
		\label{dis_bf_2}}
	\hfil
	\subfloat[$b \to \infty $]{\includegraphics[width=2in]{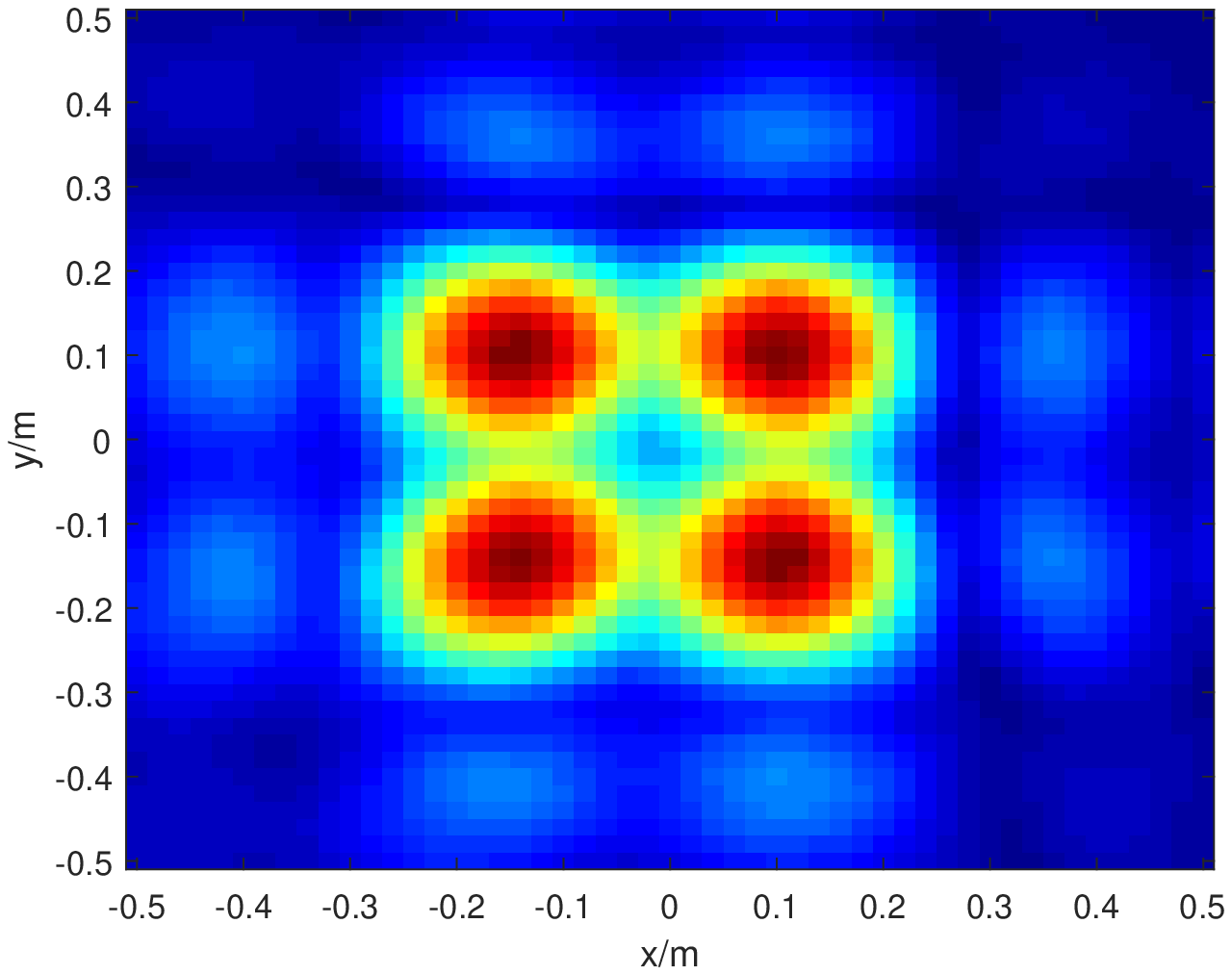}%
		\label{dis_bf_inf}}
	\hfil
	\caption{The discrete phase-shift levels versus the imaging performance of beamforming: (a) the imaging result obtained with 1-bit phase shifters, i.e., $b=1$; (b) the imaging result obtained with 2-bit phase shifters, i.e., $b=2$; (c) the imaging result obtained with continuous phase shift, i.e., $b \to \infty$.}
	\label{dis-bf}
\end{figure*}

Firstly, we evaluate the influence of discrete phase-shift levels on beamforming. Since the capital letter shaped objects could not be well reconstructed by beamforming, we still utilize the point-like target to test the impact of discrete phase-shift level on beamforming. The ground truth contains four points whose spacing is 22 cm, and their Cartesian coordinates are (-0.12,-0.12), (-0.12,0.1), (0.1,-0.12) and (0.1,0.1), respectively. The RIS with $13 \times 13 $ elements is utilized. Fig.~\ref{dis_bf_1}, Fig.~\ref{dis_bf_2}, and Fig.~\ref{dis_bf_inf} show the imaging results of the point-like target with $b = 1$, $b = 2$, and $b \to \infty $, respectively. We can observe that using the RIS with fewer phase-shift levels, i.e., smaller $b$, suffers performance loss. Especially, when the number of control bits decreases from 2 to 1, the imaging performance degrades significantly. 

\begin{figure*}[htbp]
 	\centering	
 	\subfloat[$b=1$]{\includegraphics[width=2in]{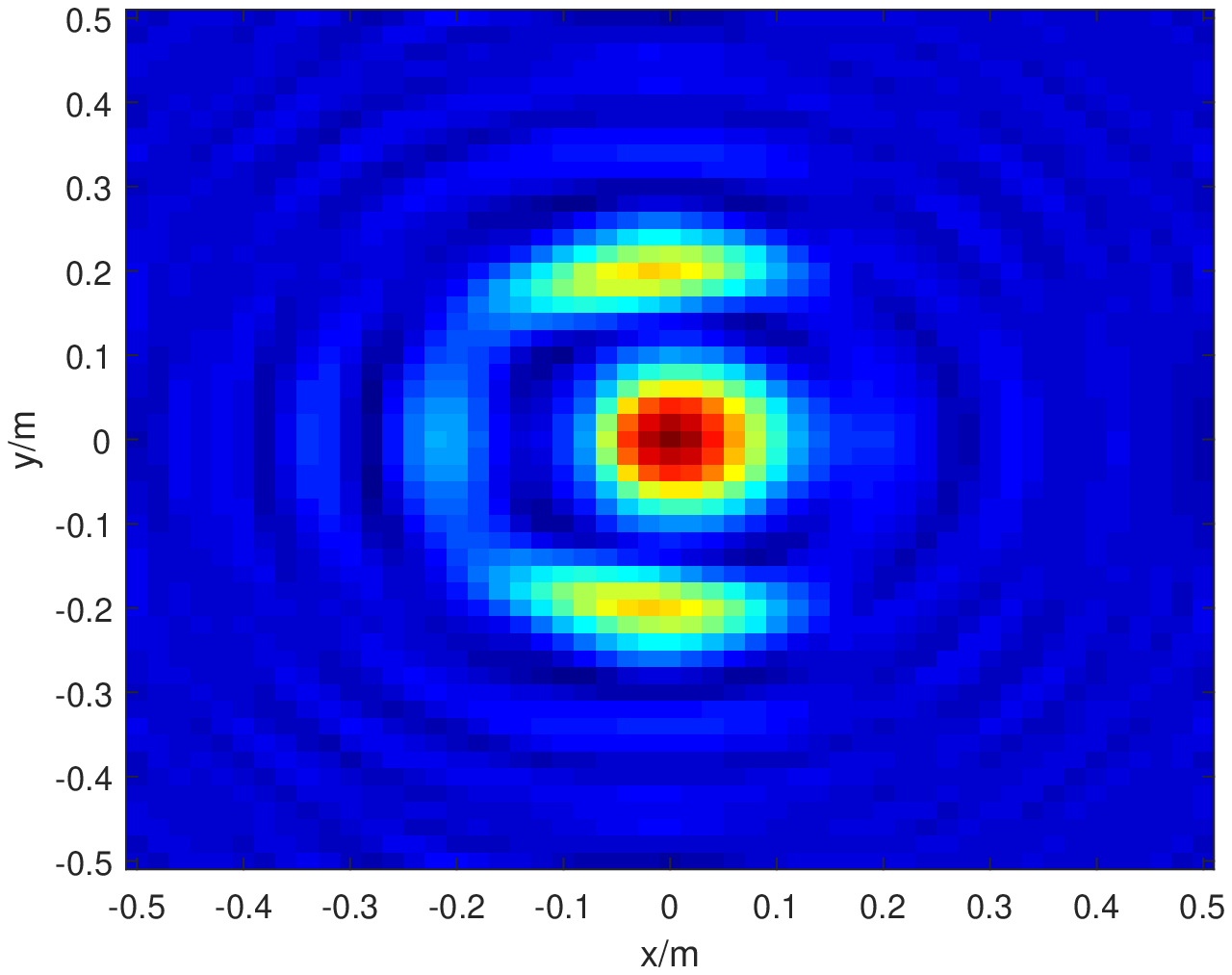}%
 		\label{dis_op_1}}
 	\hfil
 	\subfloat[$b=2$]{\includegraphics[width=2in]{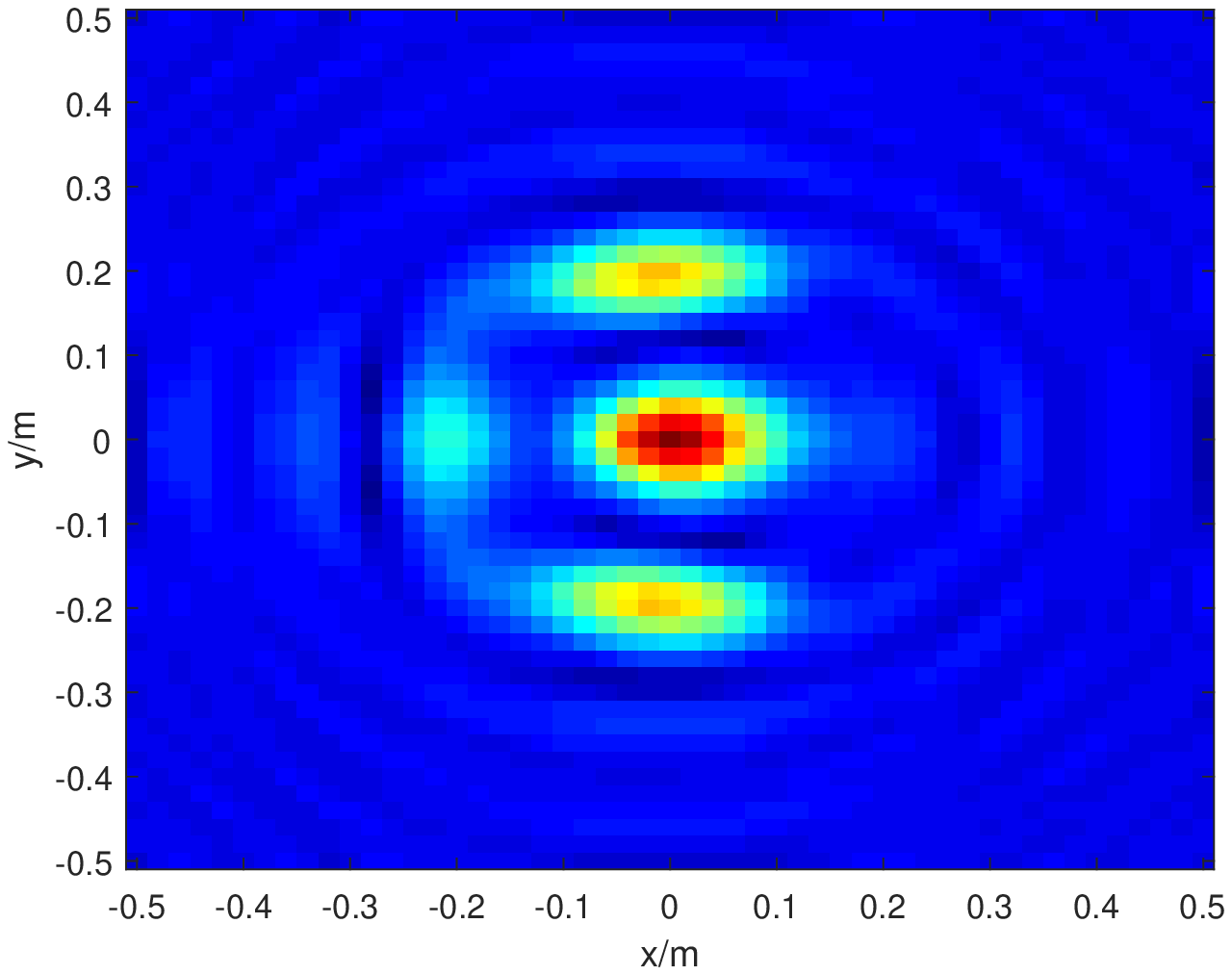}%
 		\label{dis_op_2}}
 	\hfil
 	\subfloat[$b \to \infty $]{\includegraphics[width=2in]{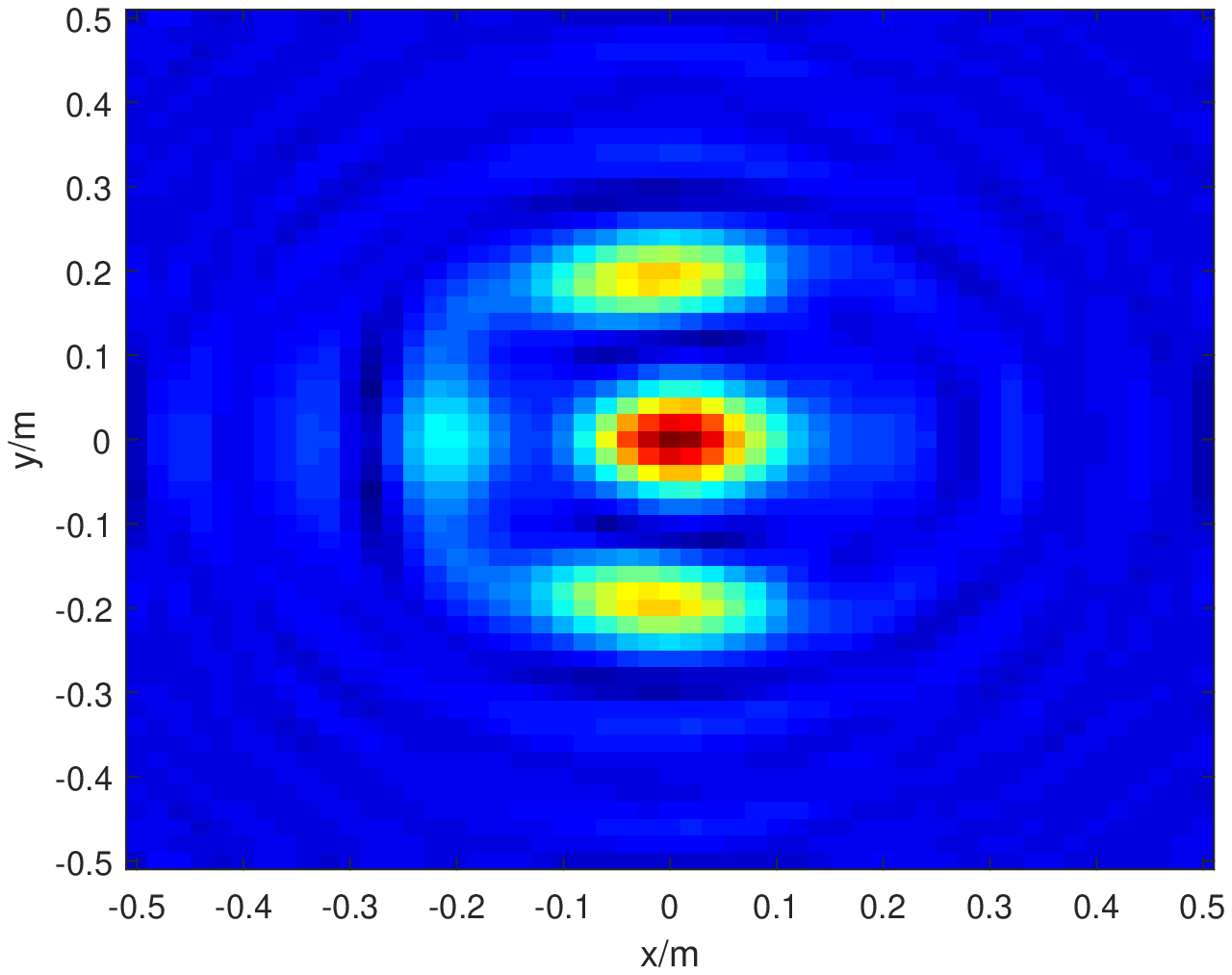}%
 		\label{dis_op_inf}}
 	\caption{The discrete phase-shift levels versus the imaging performance of optimization: (a) the imaging result obtained with 1-bit phase shifter, i.e., $b=1$; (b) the imaging result obtained with 2-bit phase shifter, i.e., $b=2$; (c) the imaging result obtained with continuous phase shift, i.e., $b \to \infty$.}
 	\label{dis-op}
 \end{figure*}

Then, we evaluate the impact of discrete phase-shift levels on the optimization-based imaging. We leverage an RIS with $7\times 7$ elements to reconstruct an E-shaped object. The letter ``E" with size of $ 50 \times 50 cm$, and the width of each line in the letter ``E" is 10 cm.  We can observe from Fig.~\ref{dis_op_1}, Fig.~\ref{dis_op_2}, and Fig.~\ref{dis_op_inf} that the performance of the optimization-based imaging is more robust to the phase-shift levels. While the phase-shift level changes from continuous to binary, the imaging results just suffer a little degradation. This is because the phase shift error is considered in the optimization processing through the matrix $\boldsymbol{H}$. As show in (13), the phase shift error caused by approximating the phase-shift values to their nearest values in the discrete sets lead to inaccurate beamforming result $\boldsymbol{p}$. On the other hand, the result of optimization-based imaging is obtained by solving (16). In such a case, if we compute the element of $\boldsymbol{H}$ with the discrete phase shift, we can still obtain an accurate estimation of the albedo $\boldsymbol{v}$. 

\begin{figure*}[htbp]    
	\centering
	\subfloat[Gound Truth]{\includegraphics[width=1.5in]{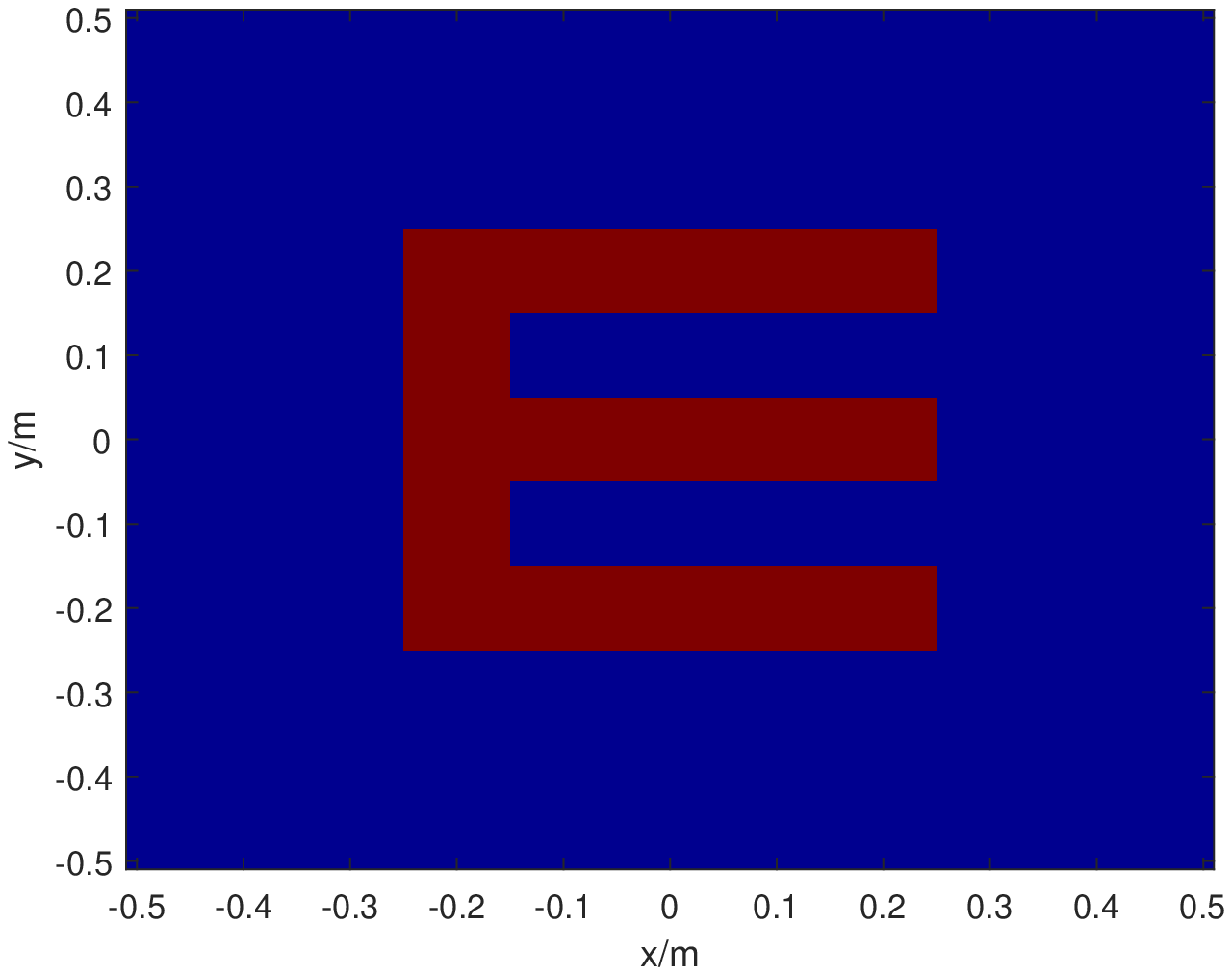}%
		\label{E1}}
	\hfil
	\subfloat[RIS: $7\times7$]{\includegraphics[width=1.5in]{Dis_1b_ob1_mine.eps}%
		\label{E2}}
	\hfil
	\subfloat[RIS: $9\times9$]{\includegraphics[width=1.5in]{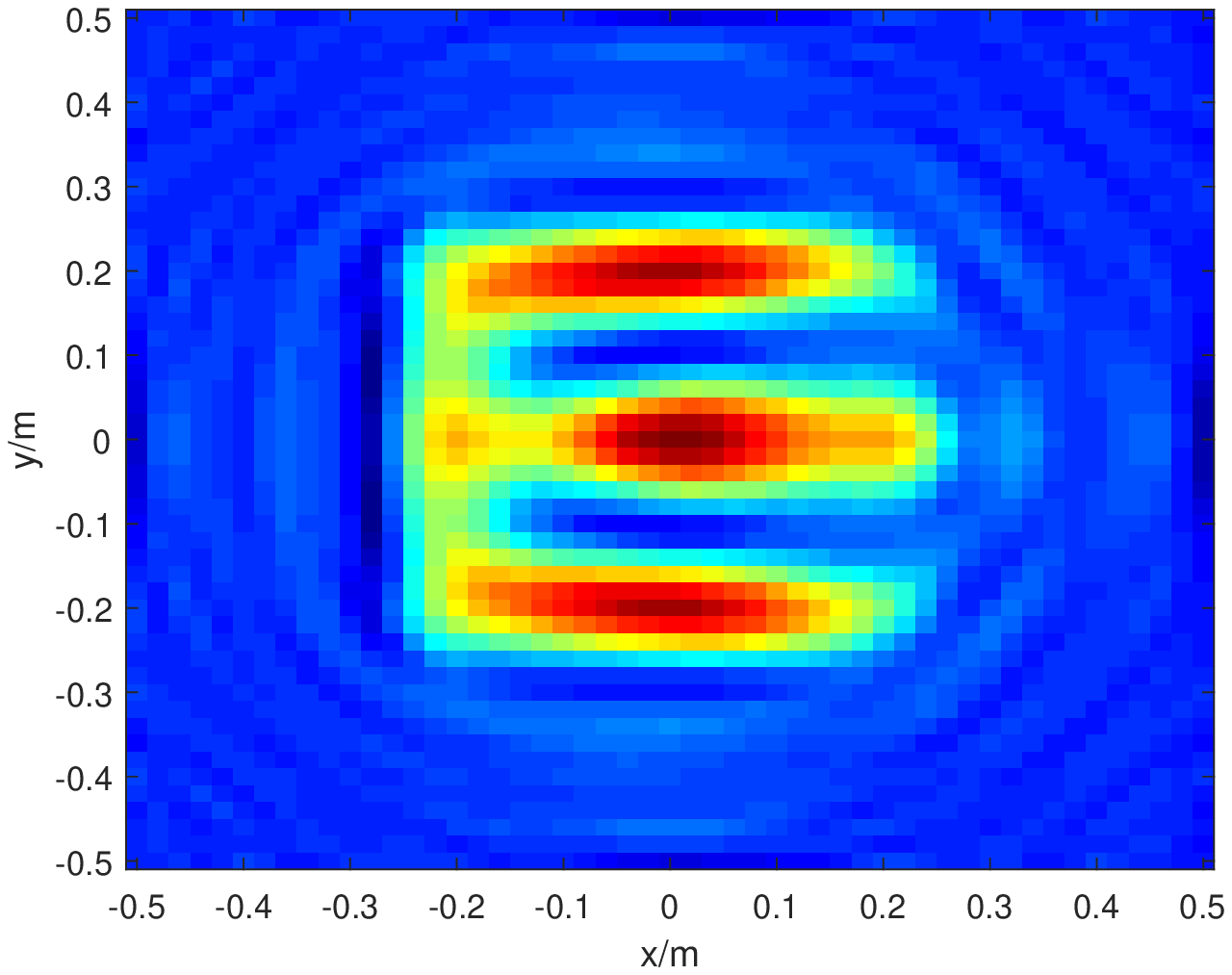}%
		\label{E3}}
	\hfil
	\subfloat[RIS: $11\times11$]{\includegraphics[width=1.5in]{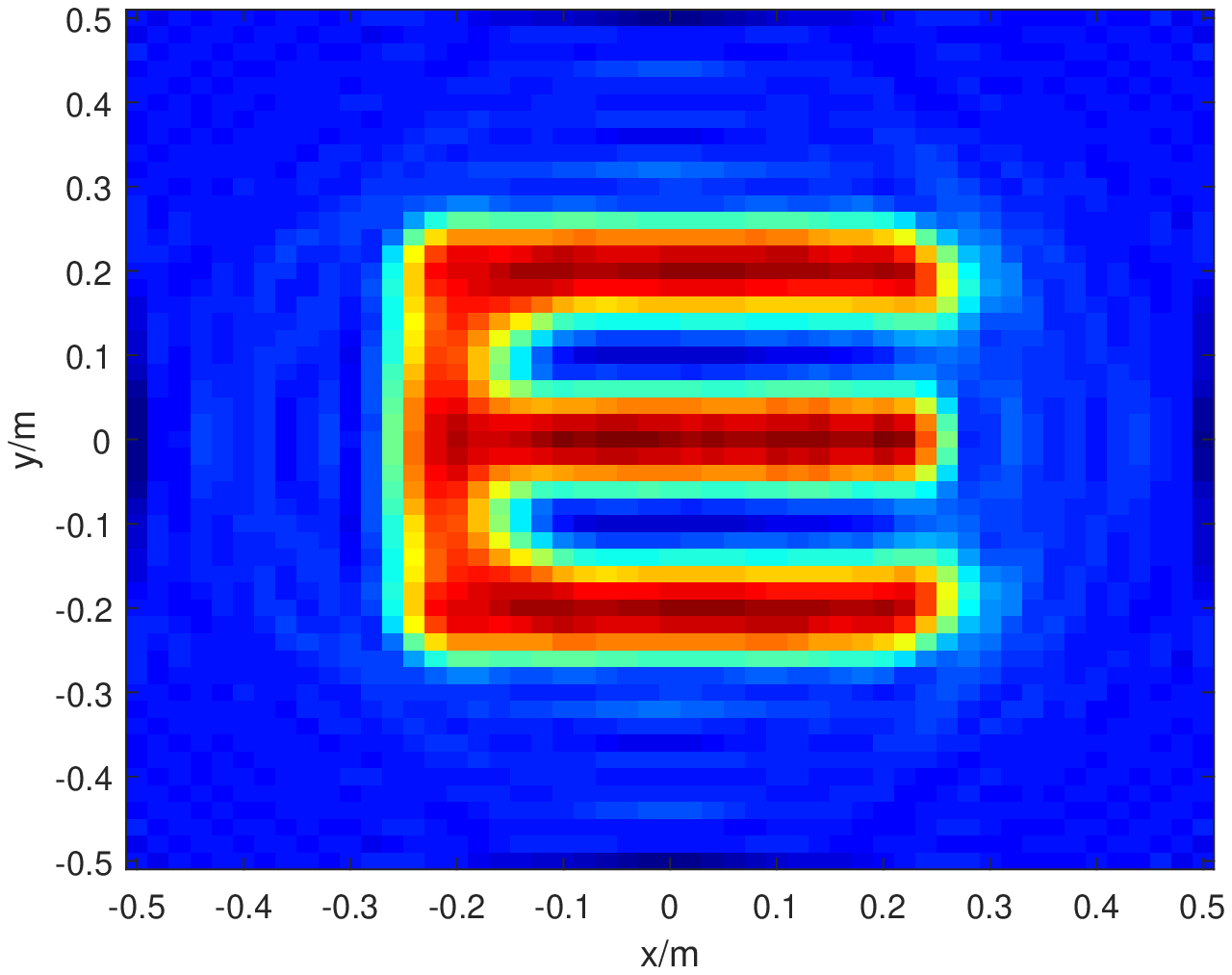}%
		\label{E4}}
	\caption{The number of RIS elements verse the imaging performance of optimization-based method when 1-bit phase shifter is used: (a) RIS with $7\times7$ elements; (b) RIS with with $9\times9$ elements; (c) RIS with $11\times11$ elements.}
	\label{fig_sim}
\end{figure*}

Fig.~\ref{E2} shows the imaging result of a letter ``E" where the RIS contains $7\times7$ elements with 1-bit phase shift. When the phase-shift level is binary, i.e., the phase shift is either 0 or "$\pi$", the imaging performance of the optimization-based algorithm degrades severely. Nevertheless, the performance can be improved by increasing the number of RIS elements. The imaging results of with a RIS of $9\times9$ and $11\times11$ are shown in Fig.~\ref{E3} and Fig.~\ref{E4}, respectively. We can see that when the number of elements increases from $7\times7$ to $11\times11$, the imaging performance becomes better. Therefore, it is very important to make a balance between the phase-shift level and the number of RIS elements according to the required imaging performance.

\section{Conclusion}
It is known that the resolution of WiFi imaging is poor due to the fundamental limitation of the bandwidth and the number of antennas of the off-the-shelf WiFi devices. In this paper, we proposed a high-resolution WiFi imaging framework with the aid of RIS to overcome the resolution limitation of traditional WiFi imaging system. By controlling the phase shift of the RIS elements, we first performed beamforming at the receiver to separate the signals coming from different spatial locations. Then, an optimization-based super-resolution imaging algorithm was proposed to achieve high-resolution reconstruction. We also considered the effect of discrete phase-shift level on the imaging performance. Finally, extensive simulation results were conducted to show that the proposed framework could indeed well reconstruct the target even with binary phase-shift level when the number of RIS elements is reasonably large, e.g., $13\times13$. We believe that such a framework could enable many applications in the near future RIS-aided WiFi/6G networks.

\bibliographystyle{IEEEtran}
\bibliography{ref2}

\end{document}